\documentclass[useAMS,usenatbib,usegraphicx]{mn2e}

\newcommand{\A}[1]{\bmath{#1}}
\newcommand{\sign}{\textrm{sign}}
\newcommand{\pd}[2]{\frac{\partial #1}{\partial #2}}
\newcommand{\DS}{\displaystyle}

\newcommand{\HALF}{\frac{1}{2}}

\newcommand{\qed}{\nobreak \ifvmode \relax \else
      \ifdim\lastskip<1.5em \hskip-\lastskip
      \hskip1.5em plus0em minus0.5em \fi \nobreak
      \vrule height0.75em width0.5em depth0.25em\fi}

\title[An HLLC Solver for Relativistic Flows]
      {An HLLC Solver for Relativistic Flows -- II. Magnetohydrodynamics}

\author[A. Mignone and G. Bodo]{A. Mignone$^{1}$\thanks{E-mail:
mignone@to.astro.it}        and G. Bodo$^{1}$\\
$^{1}$INAF Osservatorio Astronomico di Torino, 10025 Pino Torinese}
\begin{document}

\date{Accepted ??. Received ??; in original form ??}

\pagerange{\pageref{firstpage}--\pageref{lastpage}} \pubyear{2005}

\maketitle

\label{firstpage}

\begin{abstract}
An approximate Riemann solver for the equations of relativistic
magnetohydrodynamics (RMHD) is derived. The HLLC solver, originally 
developed by  Toro, Spruce and Spears, generalizes 
the algorithm described in a previous paper (Mignone \& Bodo 2004) to the 
case where magnetic fields are present.
The solution to the Riemann problem is approximated by two constant states 
bounded by two fast shocks and separated by a tangential wave.
The scheme is Jacobian-free, in the sense that it avoids the expensive 
characteristic decomposition of the RMHD equations and it improves 
over the HLL scheme by restoring the missing contact wave.

Multidimensional integration proceeds via the single step, corner transport
upwind (CTU) method of Colella, combined with the contrained tranport (CT)
algorithm to preserve divergence-free magnetic fields.  
The resulting numerical scheme is simple to implement, 
efficient and suitable for a general equation of state. 
The robustness of the new algorithm is validated against one and two 
dimensional numerical test problems.
\end{abstract}


\section{Introduction}
%
%
%
%
%

Strong evidence nowadays supports the general idea that relativistic plasmas 
may be closely related with most of the violent phenomena observed in astrophysics.
Most of these scenarios are commonly believed to involve strongly 
magnetized plasmas around compact objects.
Accretion onto super-massive black holes, for example, is invoked 
as the primary mechanism to power highly energetic phenomena observed 
in active galactic nuclei, \citep{Macchetto99,ERZ02,McK05,Shapiro05}.
In this respect, the formation and propagation of relativistic jets and 
the accretion flow dynamics pose some of the most challenging and interesting 
quests in modern theoretical astrophysics. 
Likewise, a great deal of attention has been addressed, in the last years, 
to the darkling problem of gamma ray bursts 
\citep[see for example][]{MR94,McFW99,KG02,RRD03}, 
whose models often appeal to strongly relativistic collimated outflows 
\citep{Aloy_etal00, Aloy_etal02}.
Other attractive examples include pulsar wind nebulae \citep{BZAV05},
microquasars \citep{Meier03,McKG04}, X-ray binaries \citep{VRT02} and 
stellar core collapse in the context of general relativity
\citep{Bruenn85, DFM02}.

Theoretical investigations based on direct numerical simulations 
have paved a way towards a better understanding of the rich
phenomenology of relativistic magnetized plasmas.
Part of this accomplishment owes to the successful 
generalization of existing shock-capturing Godunov-type codes to 
relativistic magnetohydrodynamics (RMHD) 
\citep[see][and reference therein]{K99, Balsara01,dZBL03}.
Implementation of such codes is based on a conservative formulation
which requires an exact or approximate solution to the 
Riemann problem, i.e., the decay of a discontinuity separating
two constant states \citep{Toro97}.
In terms of computational cost, employment of exact relativistic Riemann 
solvers may become prohibitive due to the high
degree of intrinsic nonlinearity present in the equations.
This has focused most computational efforts towards the development 
of approximate solvers which, nevertheless, require
knowledge of the exact solution, at least on some level
\citep{MM03}.
The presence of magnetic fields further entangles the solution,
since the number of decaying waves increases from 
three to seven \citep{AP87,Anile89}. An exact analytical approach 
to the solution (which does not allow compound waves) has been 
recently presented in \cite{GR05}, while 
\cite{RMPIM05} derived a special case where the velocity and magnetic
field are orthogonal.

The trade-off between efficiency, accuracy and robustness of such 
approximate methods is still a matter of research.
Solvers based on local linearization have been presented in 
\cite{K99} (KO henceforth), \cite{Balsara01} (BA henceforth)
and \cite{KKU02}.
Despite the higher accuracy in reproducing the full wave structure, 
these solvers rely on rather expensive
characteristic decompositions of the Jacobian matrix.
Conversely, the characteristic-free formulation of Harten-Lax-van Leer  
(HLL) of \cite{HLL83} has gained increasing popularity due to 
its ease of implementation and robustness.
The HLL approach has been successfully applied to the RMHD 
equations by \citealt{dZBL03} (dZBL henceforth) as well as  
to the general relativistic case \citep[see for example][]{GmKT03,DLSS05}
and to the investigation of extragalactic jets, see \cite{LAAM05}.

Besides the computational efficiency, however, the HLL formulation 
averages the full solution to the Riemann problem into a single state,
and thus lacks the ability to resolve single intermediate waves such as Alfv{\'e}n,
contact and slow discontinuities.
In \cite{MB05} (paper I henceforth) we proposed an approach that 
cured this deficiency by restoring the missing contact wave.
The resulting scheme generalized the HLLC approximate Riemann solver 
by \citet{TSS94} to the equations of relativistic hydrodynamics
without magnetic fields.
Here, along the same lines, we propose an extension
of the HLLC solver to the relativistic magnetized case.
Similar work has been presented in the context of classical MHD 
by \cite{Gurski04} and \cite{Li05}.

The new HLLC Riemann solver is implemented in the framework of 
the corner transport upwind (CTU) method of \cite{Colella90}, 
coupled with the constrained transport (CT) evolution \citep{EH88} 
of magnetic field.
The algorithm naturally preserves the divergence-free condition 
to machine accuracy and is stable up to Courant number of $1$. 

The paper is organized as follows.
The relevant equations are given in \S\ref{sec:equations}.
In \S\ref{sec:hllc} we derive the new HLLC Riemann solver. 
Numerical tests, together with the implementation of the 
CTU-CT method are shown in \S\ref{sec:test}.

\section{The RMHD Equations}\label{sec:equations}
%
%
%
%
%

The motion of an ideal relativistic magnetized fluid is 
described by conservation of mass,
\begin{equation}\label{eq:mass}
 \partial_\alpha (\rho u^\alpha) = 0 \;,
\end{equation}
energy-momentum,
\begin{equation}\label{eq:mom}
 \partial_\alpha\Big[(\rho h + |b|^2)u^\alpha u^\beta - b^\alpha b^\beta + p\eta^{\alpha\beta}\Big] = 0 \;,
\end{equation}
and by Maxwell's equations:
\begin{equation}\label{eq:maxwell}
 \partial_\alpha(u^\alpha b^\beta - u^\beta b^\alpha) = 0 \;.
\end{equation}
see, for example, \cite{AP87} or \cite{Anile89}.
In equations (\ref{eq:mass}), (\ref{eq:mom}) and (\ref{eq:maxwell}) we have
introduced the rest mass density of the fluid $\rho$, the four velocity 
$u^\alpha$, the covariant magnetic field $b^\alpha$ and the relativistic specific  
enthalpy $h$.
The total pressure $p$ results from the sum of thermal (gas) 
pressure $p_g$ and magnetic pressure $|b|^2/2$, i.e., $p = p_g + |b|^2/2$.
In what follows we assume a flat metric, so that 
$\eta^{\alpha\beta}= \textrm{diag}(-1,1,1,1)$ is the Minkowski metric tensor.
Greek indexes run from $0$ to $3$ and are customary for covariant 
expressions involving four-vectors. Latin 
indexes (from $1$ to $3$) describe three-dimensional vectors
and are used indifferently as subscripts or superscripts.

The four-vectors $u^\alpha$ and $b^\alpha$ are related to the spatial 
components of the velocity $\A{v} \equiv (v_x, v_y, v_z)$ and 
laboratory magnetic field $\A{B} \equiv (B_x, B_y, B_z)$ through
\begin{equation}\label{eq:four-vectors} \begin{array}{ccc}
 u^\alpha & = & \DS \gamma\big(1,\; \A{v}\big)  \;,\\  \noalign{\medskip}
 b^\alpha & = & \DS \gamma\left(\A{v}\cdot\A{B} \,, \quad 
 \frac{\A{B}}{\gamma^2} + \A{v}(\A{v}\cdot\A{B})\right)\;,
\end{array} 
\end{equation}
with the normalizations 
\begin{equation}
  u^\alpha u_\alpha = -1 \;,\quad
  u^\alpha b_\alpha = 0  \;,
\end{equation}
\begin{equation}
  |b|^2 \equiv b^\alpha b_\alpha = \frac{|\A{B}|^2}{\gamma^2} + (\A{v}\cdot\A{B})^2 \;,
\end{equation}
where $\gamma = (1 - \A{v}\cdot\A{v})^{-1/2}$ is the Lorentz factor. 
We follow the same conventions used in paper I, where velocities are 
given in units of the speed of light.

Writing the spatial and temporal components of equation (\ref{eq:maxwell}) 
in terms of the laboratory magnetic field yields 
\begin{equation}\label{eq:induction}
  \pd{\A{B}}{t} =  \nabla\times(\A{v}\times\A{B}) \;, 
\end{equation}
\begin{equation}\label{eq:divB}
 \nabla\cdot\A{B}  =  0  \; ,
\end{equation}
i.e., they reduce to the familiar induction equation and the 
solenoidal condition.

For computational purposes, equations  (\ref{eq:mass})--(\ref{eq:maxwell}) 
are more conveniently put in the standard conservation form
\begin{equation}\label{eq:rmhd_eq}
 \pd{\A{U}}{t} + \sum_k \pd{\A{F}^k(\A{U})}{x^k} = 0 \; ,
\end{equation}
together with the divergence-free constraint (\ref{eq:divB}), where 
$\A{U} = (D, m_x, m_y, m_z, B_x, B_y, B_z, E)$
is the vector of conservative variables and $\A{F}^k$ are
the fluxes along the $x^k\equiv (x,y,z)$ directions.
The components of $\A{U}$ are, respectively, the laboratory 
density $D$, the three components of momentum $m_k$ and 
magnetic field $B_k$ and the total energy density $E$.
From equations (\ref{eq:mass}), (\ref{eq:mom}) and the 
definitions (\ref{eq:four-vectors}) one has
\begin{eqnarray} 
 D   & = & \rho\gamma   \;,  \label{eq:cons_var_D}\\ \noalign{\medskip}
 m_k & = & (\rho h\gamma^2 + \A{B}^2)v_k - (\A{v}\cdot\A{B})B_k  \; , 
          \label{eq:cons_var_m}\\ \noalign{\medskip}
 E   & = & \DS \rho h\gamma^2  - p_g  
            + \frac{\A{B}^2}{2} + \frac{\A{v}^2\A{B}^2 - (\A{v}\cdot\A{B})^2}{2}
             \label{eq:cons_var_E}\;,
\end{eqnarray}
and
\begin{equation}\label{eq:fluxes}
  \A{F}^x(\A{U}) = \left(\begin{array}{c}
  Dv_x                    \\  \noalign{\medskip}
  \DS m_xv_x - B_x\frac{b_x}{\gamma} + p  \\  \noalign{\medskip}
  \DS m_yv_x - B_x\frac{b_y}{\gamma}  \\  \noalign{\medskip}
  \DS m_zv_x - B_x\frac{b_z}{\gamma}  \\  \noalign{\medskip}
     0         \\  \noalign{\medskip}
  B_yv_x - B_xv_y         \\  \noalign{\medskip}
  B_zv_x - B_xv_z         \\  \noalign{\medskip}
  m_x \end{array}\right)  \;.
\end{equation}
Similar expressions hold for $\A{F}^y(\A{U})$ and $\A{F}^z(\A{U})$ by 
cyclic permutations of the indexes.
Notice that the fluxes entering in the induction equation
are the components of the electric field which, in the infinite
conductivity approximation, becomes
\begin{equation}
 \A{\Omega} = -\A{v}\times\A{B} \;.
\end{equation}
The non-magnetic case is recovered by letting $\A{B}\to 0$ in
the previous expressions.

Finally, proper closure is provided by specifying an additional 
equation of state. Throughout the following we will 
assume a constant $\Gamma$-law, with specific enthalpy given
by
\begin{equation}\label{eq:eos}
  h = 1 + \frac{\Gamma}{\Gamma - 1}\frac{p_g}{\rho} \;,
\end{equation}
where $\Gamma$ is the constant specific heat ratio.

\subsection{Recovering primitive variables}
\label{sec:contoprim}
%
%

Godunov-type codes are based on a conservative formulation
where laboratory density, momentum, energy and magnetic fields 
are evolved in time.
On the other hand, primitive variables, $\A{V} = (\rho, \A{v}, p_g, \A{B})$, 
are required when computing the fluxes 
(\ref{eq:fluxes}) and more convenient for interpolation
purposes.

Recovering $\A{V}$ from $\A{U}$ is not a straightforward task 
in RMHD and different approaches have been suggested by previous authors:
BA used an iterative scheme based on a $5\times 5$ Jacobian sub-block 
of the system (\ref{eq:rmhd_eq}); KO solves a $3\times 3$ nonlinear system
of equations; dZBL (the same approach is also used in 
\cite{LAAM05}) further reduced the problem to a $2\times 2$
system of nonlinear equations.
Here we reduce this task to the solution of a single nonlinear equation, 
by properly choosing the independent variable.
If one sets, in fact, $W = \rho h \gamma^2$, $S = \A{m}\cdot\A{B}$, the 
following two relations hold:
\begin{equation}\label{eq:c2p_E}
E = W - p_g + \left(1 - \frac{1}{2{\gamma}^2}\right)|\A{B}|^2
            -\frac{S^2}{2 W^2}  \;,
\end{equation}
\begin{equation}\label{eq:c2p_m}
|\A{m}|^2 =  \left(W + |\A{B}|^2\right)^2 
           \left(1 - \frac{1}{\gamma^2}\right) -
           \frac{S^2}{W^2} \left(2W + |\A{B}|^2\right) \;.
\end{equation}

Since at the beginning of each time step $\A{m}$, $\A{B}$ and $S$ are
known quantities, equation (\ref{eq:c2p_m}) allows one to express the 
Lorentz factor $\gamma$ as a function of $W$ alone:
\begin{equation}\label{eq:lorentz}
 \gamma = \left(1 - \frac{S^2(2W + |\A{B}|^2) + |\A{m}|^2W^2}{(W + |\A{B}|^2)^2W^2}
          \right)^{-\HALF}\;.
\end{equation}

Using the equation of state (\ref{eq:eos}), the thermal pressure 
$p_g$ is also a function of $W$:
\begin{equation}\label{eq:c2p_eos}
p_g(W) = \frac{W - D\gamma}{\Gamma_r\gamma^2} \;,
\end{equation}
where $\Gamma_r = \Gamma/(\Gamma - 1)$ and $\gamma$ is now given 
by (\ref{eq:lorentz}).
Thus the only unknown appearing in equation (\ref{eq:c2p_E}) is 
$W$ and 
\begin{equation}\label{eq:press_fun}
f(W) \equiv W - p_g + \left(1 - \frac{1}{2{\gamma}^2}\right)|\A{B}|^2
            -\frac{S^2}{2 W^2} - E  = 0 \;
\end{equation}
can be solved by any standard root finding algorithm.
Although both the secant and Newton-Raphson methods have 
been implemented in our numerical code, we found the latter to be more robust and 
computationally efficient and it will be our method of choice.
The expression for the derivative needed in the Newton scheme 
is computed as follows:
\begin{equation}
 \frac{df(W)}{dW} = 1 - \frac{dp_g}{dW} + 
                    \frac{|\A{B}|^2}{\gamma^3} \frac{d\gamma}{dW} +
                    \frac{S^2}{W^3}  \;,
\end{equation}
where $dp_g/dW$ is computed from (\ref{eq:c2p_eos}), whereas 
$d\gamma/dW$ is computed from eq. (\ref{eq:lorentz}):
\begin{equation} \begin{array}{c}
 \DS \frac{dp_g}{dW} = \frac{\gamma(1 + Dd\gamma/dW) - 2Wd\gamma/dW}
                          {\Gamma_r\gamma^3} \;, \\ \noalign{\bigskip}
 \DS \frac{d\gamma}{dW} = -\gamma^3\,\frac{2S^2(3W^2 + 3W|\A{B}|^2 + |\A{B}|^4) +
                    |\A{m}|^2W^3}{2W^3(W + |\A{B}|^2)^3} \;.
\end{array}\end{equation}

Once $W$ has been computed to some accuracy, the Lorentz 
factor can be easily found from (\ref{eq:lorentz}), 
thermal pressure from (\ref{eq:c2p_eos}) and velocities 
are found by inverting equation (\ref{eq:cons_var_m}): 
\begin{equation}
  v_k = \frac{1}{W + |\A{B}|^2}\left(m_k + \frac{S}{W}B_k\right)
\end{equation}

Finally, equation (\ref{eq:cons_var_D}) is used to determine
the proper density $\rho$.

\subsection{The Riemann Problem in RMHD}\label{sec:riemann}
%
%

In the standard Godunov-type formalism, numerical integration of 
(\ref{eq:rmhd_eq}) depends on the computation of numerical fluxes 
at zone interfaces. This task is accomplished by the 
(exact or approximate) solution of the initial value problem:
\begin{equation}\label{eq:riemann}
 \A{U}(x,0) = \left\{\begin{array}{ccc}
   \A{U}_{L,i+\HALF} & \quad \textrm{if} \; & x < x_{i+\HALF} \;, \\ \noalign{\medskip}
   \A{U}_{R,i+\HALF} & \quad \textrm{if} \; & x > x_{i+\HALF} \;, \\ \noalign{\medskip}
\end{array}\right.
\end{equation}
where $\A{U}_{L,i+\HALF}$ and $\A{U}_{R,i+\HALF}$ are assumed to be
piece-wise constant left and right states at zone interface $i+\HALF$.
The evolution of the discontinuity (\ref{eq:riemann}) constitutes
the Riemann problem.

As in classical MHD, evolution in a given direction is governed 
by seven equations in seven independent conserved variables.
Integration along the $x$-direction, for example, leaves 
$B_x$ unchanged since the corresponding flux is identically zero, 
eq. (\ref{eq:fluxes}).
The solution to the initial value problem (\ref{eq:riemann}) 
results, therefore, in the formation of seven waves: two pairs 
of magneto-acoustic waves, two Alfv{\'e}n waves and a contact discontinuity. 

The complete analytical solution to the relativistic MHD Riemann
problem has been recently derived in closed form 
by \cite{GR05}. A number of properties regarding simple waves are also
well established, see \cite{AP87} and \cite{Anile89}.
\cite{RMPIM05} discuss the case in which the magnetic field of the initial 
states is tangential to the discontinuity and orthogonal to the flow velocity. 

General guidelines, relevant to the present work, follow below.
Across a magneto-acoustic (fast or slow) shock, 
all components of $\A{V}$ can change discontinuously. 
Thermodynamic quantities (e.g., $\rho$ and $p_g$) 
are continuous through a relativistic Alfv{\'e}n wave 
(as in the classical case), but contrary to the classical 
counterpart, the magnetic field is elliptically polarized and the normal 
component of the velocity is discontinuous \citep{K97}. 
Through the contact mode, only density exhibits a jump while
thermal pressure, velocity and magnetic field are continuous.

For the special case in which the component of the magnetic field 
normal to a zone interface vanishes, a degeneracy occurs where tangential,
Alfv{\'e}n and slow waves all propagate at the speed of the
fluid and the solution simplifies to a three-wave pattern.
Under this condition, the approximate solution outlined 
in paper I can still be applied with minor modifications, 
see \S\ref{sec:bx0} in this paper and \cite{MMB05}.

\section{The HLLC Solver}\label{sec:hllc}
%
%
%
%
%

The derivation of the HLL and HLLC approximate Riemann solvers has 
already been discussed in paper I and will not be repeated hereafter.

Following the same notations, we approximate the 
solution to the initial value problem (\ref{eq:riemann}) with 
two constant states, $\A{U}^*_L$ and $\A{U}^*_R$, bounded by two fast 
shocks and a contact discontinuity in the middle.
We write the solution on the $x/t=0$ axis as
\begin{equation}\label{eq:hllc_states}
  \A{U}(0,t) = \left\{\begin{array}{ccc}
   \A{U}_L    & \quad \textrm{if} & \; \lambda_L \ge 0    \;,             \\ \noalign{\medskip}
   \A{U}^*_L  & \quad \textrm{if} & \; \lambda_L \le 0 \le \lambda^* \;,\\ \noalign{\medskip}
   \A{U}^*_R  & \quad \textrm{if} & \; \lambda^* \le 0 \le \lambda_R \;,\\ \noalign{\medskip}
   \A{U}_R    & \quad \textrm{if} & \; \lambda_R \le 0              \;, \\ \noalign{\medskip}
\end{array}\right. 
\end{equation}
where $\lambda_L$ and $\lambda_R$ are, respectively, the minimum and maximum characteristic
signal velocities and $\lambda^*$ is the velocity of the middle contact wave.
The corresponding inter-cell numerical fluxes are: 
\begin{equation}\label{eq:hllc_flux}
 \A{f} = \left\{\begin{array}{ccc}
   \A{F}_L    & \quad \textrm{if} & \; \lambda_L \ge 0              \;, \\ \noalign{\medskip}
   \A{F}^*_L  & \quad \textrm{if} & \; \lambda_L \le 0 \le \lambda^*\;, \\ \noalign{\medskip}
   \A{F}^*_R  & \quad \textrm{if} & \; \lambda^* \le 0 \le \lambda_R\;, \\ \noalign{\medskip}
   \A{F}_R    & \quad \textrm{if} & \; \lambda_R \le 0              \;. \\ \noalign{\medskip}
\end{array}\right. 
\end{equation}

The intermediate fluxes $\A{F}^*_L$ and $\A{F}^*_R$ are expressed
in terms of $\A{U}^*_L$ and $\A{U}^*_R$ through  the 
Rankine-Hugoniot jump conditions:
\begin{equation}\label{eq:jump_1}\begin{array}{ccc}
  \lambda_L \left(\A{U}^*_L - \A{U}_L\right)   & = & \A{F}^*_L - \A{F}_L   \;,\\ \noalign{\medskip}
  \lambda^* \left(\A{U}^*_R - \A{U}^*_L\right) & = & \A{F}^*_R - \A{F}^*_L \;,\\ \noalign{\medskip}
  \lambda_R \left(\A{U}_R   - \A{U}^*_R\right) & = & \A{F}_R   - \A{F}^*_R \;,\\ \noalign{\medskip}
\end{array}\end{equation}
where, in general, $\A{F}^*_{L,R} \neq \A{F}(\A{U}^*_{L,R})$. 
 
The consistency condition is obtained by adding the previous equations
together:
\begin{equation}\label{eq:consistency1}
 \frac{(\lambda^* - \lambda_L) \A{U}^*_L + 
 (\lambda_R - \lambda^*) \A{U}^*_R}{\lambda_R - \lambda_L} = \A{U}^{hll}  \;,
\end{equation}  
where  
\begin{equation}\label{eq:hll_state}
 \A{U}^{hll} = \frac{\lambda_R \A{U}_R - \lambda_L\A{U}_L +
                      \A{F}_L - \A{F}_R}{\lambda_R - \lambda_L} \,,
\end{equation}
is the \emph{state} integral average of the solution to the Riemann
problem.

Similarly, if one divides each expression in eq. (\ref{eq:jump_1}) by the 
corresponding $\lambda$'s on the left hand sides and adds the resulting 
expressions, 
\begin{equation}\label{eq:consistency2}
 \frac{\A{F}^*_L\lambda_R(\lambda^* - \lambda_L) + 
   \A{F}^*_R\lambda_L(\lambda_R - \lambda^*)}{\lambda_R - \lambda_L} = \lambda^*\A{F}^{hll}\;,
\end{equation}
with
\begin{equation}\label{eq:hll_flux}
 \A{F}^{hll} = \frac{\lambda_R\A{F}_L - \lambda_L\A{F}_R + \lambda_R\lambda_L 
               (\A{U}_R - \A{U}_L)}{\lambda_R - \lambda_L}   \,.
\end{equation}
being the \emph{flux} integral average of the solution to the Riemann problem.

Since the sets of jump conditions across the contact discontinuity differ
depending on whether $B_x$ vanishes or not, we proceed by separately 
discussing the two cases.
In either case, the speed of the contact wave is assumed to be equal to the 
(average) normal velocity over the Riemann fan, i.e. $\lambda^* \equiv v^*_x$.
The normal component of magnetic field, $B_x$, is assumed to be continuous at the
interface, so that $B_x^* \equiv B_{x,L} = B_{x,R}$ can be regarded 
as a parameter in the solution.

\subsection{Case $B^*_x \neq 0$}\label{sec:bx}
%
%

We start by noticing that equations (\ref{eq:consistency1}) and 
(\ref{eq:consistency2}) provide a total of 14 relations.
Six additional conditions come by imposing continuity of total pressure, 
velocity and magnetic field components across the contact discontinuity.
This gives us a freedom of $20$ independent unknowns, $10$ per state;
we choose to introduce the following set of unknowns for each state 
\begin{equation}\label{eq:hllc_vars}
  \left\{D^*, \, v_x^*,\, v_y^*,\, v_z^*,\, B_y^*,\, B_z^*,\, m_y^*,\, m_z^*,\, E^*,\, p^*\right\}\;.
\end{equation}
The normal component of momentum ($m_x^*$) is not an independent 
variable since we assume, for consistency, that
\begin{equation}\label{eq:mE_rel}
 m_x^* = (E^* + p^*)v_x^* - \left(\A{v}^*\cdot\A{B}^*\right)B_x^* \;.
\end{equation}
The previous relation obviously holds between conservative and primitive 
physical quantities.
We point out that the choice (\ref{eq:hllc_vars}) is not unique and 
alternative sets of independent variables may be adopted.
 
According to the previous definitions, the state vector solution to the 
Riemann problem is written as 
\begin{equation}\label{eq:hllc_u*}
 \A{U}^* = \Big(D^*, m_x^*, m_y^*, m_z^*, B_y^*, B_z^*, E^*\Big)^t \;.
\end{equation}
while the flux vector, eq. (\ref{eq:fluxes}), becomes
\begin{equation}\label{eq:hllc_f*}
 \A{F}^* = \left(\begin{array}{c}
   D^*v^*_x                    \\ \noalign{\medskip}
\DS    m_x^*v^*_x - \frac{B_x^*B_x^*}{(\gamma^*)^2}
  - B^*_xv^*_x \left(\A{v}^*\cdot\A{B}^*\right) + p^* \\ \noalign{\medskip}
\DS    m_y^*v^*_x - \frac{B_x^*B_y^*}{(\gamma^*)^2}
  - B^*_xv^*_y \left(\A{v}^*\cdot\A{B}^*\right)       \\ \noalign{\medskip}
\DS    m_z^*v^*_x - \frac{B_x^*B_z^*}{(\gamma^*)^2}
  - B^*_xv^*_z \left(\A{v}^*\cdot\A{B}^*\right)       \\ \noalign{\medskip}    
   B^*_yv^*_x - B^*_xv^*_y  \\ \noalign{\medskip}
   B^*_zv^*_x - B^*_xv^*_z  \\ \noalign{\medskip}
    m^*_x  
\end{array}\right)
\end{equation}
As in paper I, we adopt the convention that quantities without the
$L$ or $R$ suffix refer indifferently to the left ($L$) or 
right ($R$) state.

The six conditions across the contact discontinuity are
\begin{equation}\begin{array}{ccc}
  v^*_{x,L} = v^*_{x,R} \;, & v^*_{y,L} = v^*_{y,R}\;, & v^*_{z,L} = v^*_{z,R}  \;, \\ \noalign{\medskip}
  B^*_{y,L} = B^*_{y,R} \;, & B^*_{z,L} = B^*_{z,R}\;, & p^*_L     = p^*_R    \;.
\end{array}\end{equation}
For these quantities the suffix $L$ or $R$ is thus unnecessary.

From the transverse components of the magnetic field in the state
consistency condition (\ref{eq:consistency1}), one immediately finds that
\begin{equation}\label{eq:transv_B}
    B^*_{y} = B_{y}^{hll} \; ,\quad
    B^*_{z} = B_{z}^{hll} \; .
\end{equation}
Thus the transverse components the magnetic field are given by the HLL single 
state.  
Similarly, from the fifth and sixth components of the flux consistency condition 
(\ref{eq:consistency2}) one can express the transverse velocity through
\begin{equation}\label{eq:transv_v}
  B^*_xv^*_y = B^*_yv^*_x - F_{B_y}^{hll} \;,\quad
  B^*_xv^*_z = B^*_zv^*_x - F_{B_z}^{hll}  \;,
\end{equation}
where $F_{B_y}^{hll}$ and $F_{B_z}^{hll}$ are the $B_y$- and $B_z$- components of
the HLL flux, eq. (\ref{eq:hll_flux}).
Simple manipulations of the normal momentum and energy components in equation 
(\ref{eq:consistency1}) together with (\ref{eq:mE_rel}) yield the following 
simple expression:
\begin{equation}\label{eq:state_eq}
   E^{hll} v^*_x + p^*v^*_x - B^*_x\, \big(\A{v}^*\cdot\A{B}^*\big)
   = m_x^{hll} \;.
\end{equation}

Similar algebra on the momentum and energy components of the 
flux consistency condition (\ref{eq:consistency2}) leads to
\begin{equation}\label{eq:flux_eq}
  \Big[F^{hll}_E - B^*_x\,(\A{v}^*\cdot\A{B}^*)\Big]v^*_x - 
    \left(\frac{B^*_x}{\gamma^*}\right)^2 + p^* - F^{hll}_{m^x} =0 \;.
\end{equation}
where $1/(\gamma^*)^2 = 1 - (v_x^*)^2 - (v_y^*)^2 - (v_z^*)^2$.

Now, if one multiplies equation (\ref{eq:flux_eq}) by $v_x^*$ and subtracts 
equation (\ref{eq:state_eq}), the following quadratic equation 
may be obtained:
\begin{equation}\label{eq:quadratic}
  a(v_x^*)^2 + bv_x^* + c = 0 \;,
\end{equation}
with coefficients
\begin{equation}\begin{array}{ccl}
 a & = & F^{hll}_E - \A{B}^{hll}_\perp\cdot\A{F}^{hll}_{\A{B}_\perp} \;,\\ \noalign{\medskip}
 b & = & - F^{hll}_{m^x} - E^{hll} + \left|\A{B}_\perp^{hll}\right|^2 
         + \left|\A{F}^{hll}_{\A{B}_\perp}\right|^2 \;, \\ \noalign{\medskip}
 c & = &   m_x^{hll} - \A{B}^{hll}_\perp\cdot\A{F}^{hll}_{\A{B}_\perp} \;.
\end{array}\end{equation} 
In the previous expressions $\A{B}^{hll}_\perp \equiv (0,B^{hll}_y,B^{hll}_z)$, 
$\A{F}^{hll}_{\A{B}_\perp} \equiv (0,F^{hll}_{B_y}, F^{hll}_{B_z})$. 
Similar arguments to those presented in paper I lead to the conclusion
that only the root with the minus sign is physically admissible.

Once $v_x^*$ is known, $v_y^*$ and $v_z^*$ are readily obtained from 
(\ref{eq:transv_v}), $p^*$ is computed from (\ref{eq:flux_eq}),
while density, transverse momenta and energy are obtained using the 
Rankine-Hugoniot jump conditions across each fast wave: 
\begin{eqnarray} 
 D^* & = & \DS \frac{\lambda - v^x}{\lambda - v_x^*} D   
                   \;, \label{eq:D_jump} \\ \noalign{\medskip}
\label{eq:transv_my}
 m^*_y & = &\DS  \frac{-B^*_x\left[\frac{B^*_y}{(\gamma^*)^2} + (\A{v}^*\cdot\A{B}^*)v_y^*\right] 
                         + \lambda m_y - F_{m_y}}{\lambda - v^*_x}  
                   \;, \label{eq:my_jump} \\ \noalign{\medskip}
\label{eq:transv_mz}
 m^*_z & = &\DS  \frac{-B^*_x\left[\frac{B^*_z}{(\gamma^*)^2} + (\A{v}^*\cdot\A{B}^*)v_z^*\right] 
         + \lambda m_z - F_{m_z}}{\lambda - v^*_x}                
                   \;, \label{eq:mz_jump}  \\ \noalign{\medskip}
 E^* & = & \DS \frac{\lambda E - m_x + p^*v_x^* - (\A{v}^*\cdot\A{B}^*)B^*_x}{\lambda - v_x^*}  \;.
          \label{eq:E_jump}
\end{eqnarray}
In equations (\ref{eq:my_jump}) and (\ref{eq:mz_jump}), $F_{m_y}$ and $F_{m_z}$ are,
respectively, the $m_y$- and $m_z$- components of the flux, eq. (\ref{eq:fluxes}), evaluated at 
the left or right state. 
As in paper I, we have omitted the suffix $L$ or $R$ for clarity of exposition.

\subsection{Case $B^*_x = 0$}\label{sec:bx0}
%
%

For vanishing normal component of the magnetic field 
a degeneracy occurs where the Alfv{\'e}n waves and the two slow 
magnetosonic waves propagate at the speed of the contact discontinuity.
For this case the approximate character of the HLLC solver offers a better 
representation of the exact solution, since the Riemann fan is comprised of 
three waves only.
At the contact discontinuity, however, only the normal component
of the velocity $v_x$ and the total pressure $p$ are continuous
(KO). The remaining variables experience jumps. 
This only adds $2$ constraints
to the $14$ jump conditions, leaving a freedom of $8$ unknowns per state.
However, the transverse velocities $v_y$ and $v_z$ do not enter explicitly 
in the fluxes (\ref{eq:hllc_f*}) and the jump conditions can be written entirely
in terms of $\{D^*, v^*_x, m^*_y, m^*_z, B^*_y, B^*_z, E^*, p^*\}$,
i.e. $8$ unknowns per state.
Straightforward algebra shows that the coefficients of the quadratic
equation (\ref{eq:quadratic}) are now given by 
\begin{equation} 
 a =   F^{hll}_E               \;,\quad 
 b = - F^{hll}_{m^x} - E^{hll}  \;, \quad
 c =    m_x^{hll}     \;,
\end{equation} 
i.e., they coincide with the expressions derived in paper I. 
The root with the minus sign still represents the correct physical
solution.
Once $v_x^*$ is found, the total pressure $p^*$ 
is derived from
\begin{equation}
  p = - F^{hll}_{E}v_x^* + F^{hll}_{m_x}\;,
\end{equation}
and the normal momentum (\ref{eq:mE_rel}) becomes 
\begin{equation}
 m_x^* = (E^* + p^*)v_x^* \;.
\end{equation}
 
The remaining quantities are easily obtained
from the jump conditions:
\begin{eqnarray}
 D^*   & = & \DS \frac{\lambda - v_x}{\lambda - v_x^*} D  \;,       
         \\ \noalign{\medskip}
 m^*_{y,z} & = & \DS \frac{\lambda - v_x}{\lambda - v^*_x}\, m_{y,z}    \;,
         \\ \noalign{\medskip}
 E^*   & = & \DS \frac{\lambda E - m_x + p^*v_x^*}{\lambda - v_x^*} \;,
     \label{eq:E_jump2}    \\ \noalign{\medskip}
 B^*_{y,z} & = & \DS \frac{\lambda - v_x}{\lambda - v_x^*}\, B_{y,z}  \;.
\end{eqnarray}

\subsection{Remarks}\label{sec:remarks}
%
%

The expressions derived separately in \S\ref{sec:bx} and 
\S\ref{sec:bx0} are suitable in the $B_x\neq0$ and $B_x\to0$ cases,
respectively.
Although other degeneracies may be present (see KO for
a thorough discussion) no other modifications are necessary 
to the algorithm. 
Before testing the new solver, however, a few remarks are worth of notice:
\begin{enumerate}
\renewcommand{\theenumi}{(\arabic{enumi})}
 \item The solutions derived separately for $B_x \neq 0$ and the special 
       case $B_x = 0$ automatically satisfy the consistency 
       conditions (\ref{eq:consistency1}) and (\ref{eq:consistency2})
       by construction;

 \item In the limit of zero magnetic field, the expressions derived 
       in \S\ref{sec:bx0} reduce to those found in paper I;

 \item In the classical limit, our derivation does not coincide with the 
       approximate Riemann solvers constructed by \cite{Gurski04} or \cite{Li05}.  
       The reason for this discrepancy stems from the fact that both 
       \cite{Gurski04} and \cite{Li05} assume that transverse momenta
       and velocities are tied by the relation $m^*_{y,z} \equiv \rho^*v_{y,z}^*$. 
       Although certainly true in the exact solution, this assumption 
       reduces, in the HLLC approximate formalism, the number of unknowns from 
       $10$ to $8$ (when $B_x\neq0$) thus leaving the systems of jump conditions
       (\ref{eq:jump_1}) overdetermined.
       Should this be the case, the number of equations exceeds the number of unknowns 
       and the integral relations across the Riemann fan inevitably break down. 
       This explains the inconsistencies found in Li's and Gurski's derivations and
       further discussed in \cite{MK05}.
       
       Therefore, in the classical limit, our expressions automatically imply 
       $m^*_{y,z} \neq \rho^*v_{y,z}^*$
       and the correct expressions for the transverse velocities are still given by 
       (\ref{eq:transv_v}), whereas transverse momenta should be derived from the 
       jump conditions accordingly.
       Furthermore, contrary to Li's misconception, consistency with the jump 
       conditions requires that the magnetic field components be uniquely determined by 
       (\ref{eq:transv_B}) and no other choices are thus possible. 
       
 \item The reader might have noticed that in the limit
       of vanishing $B_x$, some of the expressions given in 
       \S\ref{sec:bx} do not reduce to the those found in \S\ref{sec:bx0}.
       This property also persists in the classical limit, see \cite{Gurski04},
       and \cite{Li05}.
       The reason for this discrepancy relies on the assumption 
       of continuity of the transverse components of magnetic field
       across the tangential wave $\lambda^*$:
       when $B_x \to 0$, a degeneracy occurs where the tangential,
       Alfv{\'e}n and slow waves all propagate at the speed of the
       fluid and the solution simplifies to a three-wave pattern.
       In the exact solution, the continuity of $B_y$ and $B_z$ across 
       the tangential wave is lost since the middle state
       bounded by the two slow waves becomes singular.  

 \item Lastly, we note that in both the classical and relativistic
       case the transverse velocities given by eq. (\ref{eq:transv_v}) 
       become ill-defined as $B_x\to 0$. However, 
       in the classical case, the terms involving $v^*_y$ 
       or $v^*_z$ in the flux definitions remain finite as 
       $B_x\to 0$. Conversely, this is not the case 
       in RMHD for arbitrary orientation of the magnetic field
       as one can see, for example, using eq. (\ref{eq:transv_my}):
       \begin{equation}
        m^*_y \sim \frac{(B_z^{hll}v_x^* - F^{hll}_{B_z} )
                         (F^{hll}_{B_y} B^{hll}_z - F_{B_z}^{hll}B^{hll}_y)}
                         {B_x(\lambda - v^*_x)} + O(1) 
       \end{equation}
      as $B_x\to 0$. 
      Fortunately, for strictly two dimensional flows (e.g. when $B_z = v_z = 0$)
      the leading order term vanishes and the singularity 
      is avoided. 
      In the general case, however, we conclude that more sophisticated 
      solvers should allow the presence of rotational discontinuities 
      in the solution to the Riemann problem.
      This has been done, for example, by \cite{MK05} in the context
      of classical MHD.       
\end{enumerate}
%

\subsection{Wave Speed Estimate}\label{sec:speeds}
%
%
%

The full characteristic decomposition of the RMHD equation (i.e. 
the eigenvalues and eigenvectors of the Jacobian matrix 
$\partial\A{F}^x/\partial\A{U}$) was extensively analyzed by 
\cite{AP87} and \cite{Anile89}. 
In the one-dimensional case the Jacobian matrix can be decomposed 
into seven eigenvectors associated with four magnetosonic
waves (fast and slow disturbances), two Alfv{\'e}n waves and one entropy
wave propagating at the fluid velocity. 
The eigenstructure is therefore similar to the classical case and 
it can be shown that the ordering of the various speeds and corresponding 
degeneracies are preserved \citep{Anile89}.
 
Since the HLLC approximate Riemann solver requires 
an estimate of the outermost waves, the right and left-going 
fast shock speeds identify the necessary characteristic velocities.
Thus we set \citep{Davis88}:
\begin{equation}\begin{array}{c}\label{eq:wavespeeds} 
  \lambda_L = \min\big(\lambda_-(\A{V}_L), \lambda_-(\A{V}_R)\big) \;,
  \\ \noalign{\medskip}
  \lambda_R = \max\big(\lambda_+(\A{V}_L), \lambda_+(\A{V}_R)\big) \;,
  \end{array}
\end{equation}
where $\lambda_{-}$ and $\lambda_+$ are the minimum and maximum 
roots of the quartic equation
\begin{equation}\label{eq:eigenspeed_eq}
 \rho h(1-c_s^2)a^4 = (1-\lambda^2) \left[(|b|^2 + \rho h c_s^2)a^2 
  - c_s^2{\cal B}^2\right]  \;,
\end{equation} 
with $a = \gamma(\lambda - v_x)$, ${\cal B} = b^x - \lambda b^0$.
In absence of magnetic field, both the (left and right-going) 
slow and fast shocks propagate at the same speed and
equation (\ref{eq:eigenspeed_eq}) reduces to the quadratic equation
(22) shown in paper I. 
When $\A{B} \neq \A{0}$, no simple analytical expression is available 
and solving (\ref{eq:eigenspeed_eq}) requires numerical or rather
cumbersome analytical approaches.
Recently, \cite{LAAM05} proposed approximate simple lower 
and upper bounds to the required eigenvalues. 
Here we choose to solve eq. (\ref{eq:eigenspeed_eq}) by means of 
analytical methods, where the quartic is reduced to a cubic 
equation which is in turn solved by standard methods.

There are special cases where it is possible to handle some
of the degeneracies more efficiently using simple analytical formulae:
\begin{itemize}
 \item for vanishing total velocity, equation (\ref{eq:eigenspeed_eq}) 
       reduces to a bi-quadratic, 
     \begin{equation} 
         (\rho h + |b|^2)\lambda^4 -
         (|b|^2 + \rho hc_s^2 + B_x^2c_s^2)\lambda^2 +
         c_s^2B_x^2 = 0
     \end{equation}
 \item for vanishing normal component of the magnetic field,
       equation (\ref{eq:eigenspeed_eq}) yields a quadratic 
       equation: 

    \begin{equation}
      a_2 \lambda^2 + a_1 \lambda + a_0 = 0
    \end{equation}
    with $a_2 = \rho h\big[c_s^2 + \gamma^2(1-c_s^2)\big] + {\cal Q}$, 
    $a_1 = -2\rho h\gamma^2v_x(1-c_s^2)$, 
    $a_0 = \rho h\big[ - c_s^2 + \gamma^2v_x^2(1-c_s^2)\big] - {\cal Q}$
    and ${\cal Q} = |b|^2 - c_s^2(\A{v}_\perp\cdot\A{B}_\perp)^2$.
\end{itemize}
For all other cases we solve the quartic equation (\ref{eq:eigenspeed_eq}).

\subsection{Positivity of the HLLC scheme}\label{sec:positivity}
%
%
%
%

The set of physically admissible conservative states, $G$, identify
all the $\A{U}$'s yielding positive thermal pressure $p_g$ and total velocity 
$|\A{v}| < 1$, according to the procedure outlined in \S\ref{sec:contoprim}.
Thus the positivity of the HLLC approximate Riemann solver requires 
that  
\begin{itemize}
 \item both left and right intermediate states $\A{U}^*_L$ and $\A{U}^*_R$ 
       belong to $G$; 
 \item the first-order scheme yields updated conservative states that 
       are in $G$.
\end{itemize}

Unfortunately, the mathematical proof of the positivity of the HLLC scheme 
presents remarkable algebraic difficulties.
In absence of the singular behavior described in \S\ref{sec:remarks}, 
investigations have been carried at the numerical level
by verifying that each intermediate state $\A{U}^*$ correspond to a primitive,
physically admissible state. In all the tests presented in this paper and several
others not discussed here, the scheme did not manifest any loss of positivity.
However, in the general three-dimensional case when $B_x,B_y,B_z\neq0$, the terms 
involving $B_x$ in the expressions for the transverse momenta may become 
arbitrarily large as $B_x\to 0$ and a loss of positivity can be experienced.

\section{Algorithm Validation}\label{sec:test}
%
%
%
%

\subsection{Corner Transport Upwind for relativistic MHD}\label{sec:ctu}
%
%

The RMHD equations (\ref{eq:rmhd_eq}) are evolved in  
a conservative, dimensionally unsplit fashion:
\begin{equation}\label{eq:update}
 \A{U}^{n+1}_{i,j} =  \A{U}^n_{i,j} + \A{\cal L}^{x,n+\HALF}_{i,j} 
                                    + \A{\cal L}^{y,n+\HALF}_{i,j} \,,
\end{equation}
where the $\A{\cal L}$'s are Godunov operators
\begin{equation}\label{eq:god_op_x}
 \A{\cal L}^{x,n+\HALF}_{i,j} =  
 - \frac{\Delta t}{\Delta x_i}
   \left(\A{f}^{x,n+\HALF}_{i+\HALF,j} - \A{f}^{x,n+\HALF}_{i-\HALF,j}\right)\,,
\end{equation}
\begin{equation}\label{eq:god_op_y}
 \A{\cal L}^{y,n+\HALF}_{i,j} =  
 - \frac{\Delta t}{\Delta y_j}
   \left(\A{f}^{y,n+\HALF}_{i,j+\HALF} - \A{f}^{y,n+\HALF}_{i,j-\HALF}\right)\,,
\end{equation}
and $\A{U}^n$ is the set of volume-averaged conservative variables 
$\A{U}^n = \Big(D, \A{m}, \bar{\A{B}}, E\Big)^n$ at time $t=t^n$.
Here $\bar{\A{B}}$ denotes the zone-averaged magnetic field.
For clarity of exposition we will omit, throughout the following,
integer-valued subscripts $(i,j)$ and retain only the half-integer notation 
to denote zone edge values.

The fluxes appearing in equations (\ref{eq:god_op_x}) and (\ref{eq:god_op_y}) 
are computed by solving, at each zone interface, a Riemann problem 
with suitable time-centered left and right input states.
For example, we obtain $\A{f}^{y,n+\HALF}_{j+\HALF}$ 
as the HLLC flux with input states given
by $\A{V}^{n+\HALF}_{j+\HALF,L}$ and $\A{V}^{n+\HALF}_{j+\HALF,R}$,
respectively.

Computation of time-centered left and right zone edge values 
proceeds using the corner transport upwind (CTU) of \cite{Colella90},
recently extended to relativistic hydrodynamics
by \cite{MPB05} and to classical MHD by \cite{GS05}.
Here we generalize the CTU approach to relativistic MHD by following a 
slightly different approach, although equivalent to the 
guidelines given in \cite{Colella90}.
For the sake of conciseness, only the essential steps will be 
described hereafter. The unfamiliar reader is referred to 
the work of \cite{Colella90}, \cite{Saltzman94} and \cite{GS05} 
for more comprehensive derivations.

In our formulation, second-order accurate left and right states
are sought in the form 
\begin{equation}\label{eq:pred_states}
\A{V}^{n+\HALF}_{i\pm\HALF,S} = \A{V}^{x,n+\HALF} 
                                 \pm \frac{\delta_x\A{V}^n}{2} \, , \quad
\A{V}^{n+\HALF}_{j\pm\HALF,S} = \A{V}^{y,n+\HALF}
                                 \pm \frac{\delta_y\A{V}^n}{2} \, , \quad
\end{equation}
where we take $S=L$ ($S=R$) with the plus (minus) sign.
The slopes $\delta_x\A{V}^n$ and $\delta_y\A{V}^n$ are 
computed at the beginning of the time step 
using, for example, the monotonized central-difference (MC) 
limiter:
\begin{equation}\label{eq:mc_lim}
  \delta_x q^n = 
   s_i\min\left(2|\Delta q^n_+|, 2|\Delta q^n_-|, 
             \frac{|q^n_{i+1} - q^n_{i-1}|}{2}\right) \,,
\end{equation}
where $q\in\A{V}$ and
\begin{equation}
  \Delta q^n_\pm = \pm\left(q^n_{i\pm1} - q^n_i\right) \,,\;
  s_i = \frac{\sign(\Delta q^n_+) + \sign(\Delta q^n_-)}{2} \;.
\end{equation}
An alternative smoother prescription is given by the harmonic mean
\citep{vLeer77}:
\begin{equation}\label{eq:vl_lim}
 \delta_x q^n = \frac{2\max\left(0, \Delta q_+\Delta q_-\right)}
                         {\Delta q_+ + \Delta q_-} \,.
\end{equation}

Equation (\ref{eq:mc_lim}) provides smaller dissipation at 
discontinuities, whereas equation (\ref{eq:vl_lim}) was found
to give less oscillatory results.
Interpolation in the $y$-direction is done in a similar manner.
Additional forms of limiting may be adopted if necessary, 
see \S\ref{sec:shockflattening} and \S\ref{sec:mdlimit}.

The cell- and time- centered values on the right hand sides 
of equations (\ref{eq:pred_states}) are computed from a Taylor 
expansion of the conservative variables, i.e.
\begin{equation}\label{eq:cent_pred_x}
 \A{U}^{x,n+\HALF} \approx 
 \A{U}^n + \frac{\Delta t}{2}\pd{\A{U}}{t} =  
 \A{U}^n - \frac{\Delta t}{2}\left(  \pd{\hat{\A{F}}^x}{x}
                                   + \pd{\A{F}^y}{y}\right) \,,
\end{equation}
\begin{equation}\label{eq:cent_pred_y}
 \A{U}^{y,n+\HALF} \approx 
 \A{U}^n + \frac{\Delta t}{2}\pd{\A{U}}{t} =  
 \A{U}^n - \frac{\Delta t}{2}\left(  \pd{\A{F}^x}{x}  
                                         + \pd{\hat{\A{F}}^y}{y}\right) \,.
\end{equation}
Following \cite{Colella90}, we approximate the spatial derivative in 
the direction normal to a zone interface (denoted with a hat) 
with the Hancock step already introduced in paper I, 
\begin{equation}\label{eq:hancock_diff}
 \pd{\hat{\A{F}}^x}{x} \approx 
 \frac{   \A{F}^x\left(\A{V}^n_{i+\HALF,L}\right) 
        - \A{F}^x\left(\A{V}^n_{i-\HALF,R}\right)}{\Delta x_i}\,,
\end{equation}
whereas the derivative in the tangential direction is
computed in an upwind fashion using a Godunov operator:
\begin{equation}\label{eq:upwind_diff}
 \Delta t\pd{\A{F}^y}{y} \approx -\A{\cal{L}}^{y,n} = 
   \frac{\Delta t}{\Delta y_j}
   \left(\A{f}^{y,n}_{j+\HALF} - \A{f}^{y,n}_{j-\HALF}\right)\,.
\end{equation}
The state $\A{U}^{y,n+\HALF}$ is obtained by similar arguments by 
interchanging the role of normal and tangential derivatives. 
We would like to point out that the Godunov operators used in the predictor
step involve left and right states computed at $t=t^n$ 
(and not at $t=t^{n+\HALF}$ as in \cite{GS05}):
\begin{equation}\label{eq:init_states}
 \A{V}^n_{i\pm\HALF,S} = \A{V}^n \pm \frac{\delta_x\A{V}^n}{2} \;, \quad
 \A{V}^n_{j\pm\HALF,S} = \A{V}^n \pm \frac{\delta_y\A{V}^n}{2} \,.
\end{equation}
This choice still makes the scheme second-order 
accurate in space and time and was found, in our experience, 
to yield a more robust algorithm.
Besides, our CTU implementation does not require a primitive variable 
formulation, thus offering ease of implementation 
in the context of relativistic hydro and MHD, where the Jacobian 
$\partial\A{F}/\partial\A{U}$ is particularly expensive to evaluate. 

Note that a total of four Riemann problems are involved in 
the single time step update (\ref{eq:update}).
It can be easily verified that for one-dimensional
flows, the corner transport upwind method outlined
above reduces to the scheme presented in paper I. 

Finally, the choice of the time step $\Delta t$ is based on the
Courant-Friederichs-Lewy (CFL) condition \citep{CFL28}:
\begin{equation}
  \Delta t = \textrm{CFL} \times \min_{i,j}\left(
   \frac{ \Delta x}{\max(|\lambda^x_L|,|\lambda^x_R|)},
   \frac{ \Delta y}{\max(|\lambda^y_L|,|\lambda^y_R|)}\right) \,,
\end{equation}
where $0<\textrm{CFL}<1$ is the Courant number and 
$|\lambda^x_{L,R}|$, $|\lambda^y_{L,R}|$
are the zone interface wave speeds computed in the $x$ and $y$
directions according to (\ref{eq:wavespeeds}).

\subsubsection{Contrained Transport Evolution of the Magnetic Field}
\label{sec:divB}
%
%
%
%

It is well known that multidimensional numerical schemes do not 
generally preserve the solenoidal condition, eq. (\ref{eq:divB}), 
unless special discretization techniques are employed.
In this respect, several approaches have been suggested 
in the context of the classical MHD equations \citep{Toth00, LdZ00} 
and some of them have been recently extended to the relativistic
case, see dZBL.
Here we adopt the constrained transport (CT)
\citep{EH88} and follow the approach of 
\cite{BS99} for its integration in Godunov-type schemes. 

In the CT approach a new staggered magnetic field variable is introduced. 
In this representation, the components of the magnetic 
field are treated as area-weighted averages on the zone 
faces to which they are orthogonal. 
Thus, $B_x$ is collocated at $(i+\HALF,j)$, whereas 
$B_y$ at $(i,j+\HALF)$. No jump is allowed in the normal component
of $\A{B}$ at a zone boundary, consistently with the 
well posedness of the Riemann problem presented in \S\ref{sec:riemann} 
and \S\ref{sec:hllc}.   
Transverse components may be discontinuous.

In this formulation, a discrete version of Stoke's theorem is
used integrate the induction equation (\ref{eq:induction}).
For example, after the predictor steps (\ref{eq:cent_pred_x}) 
and (\ref{eq:cent_pred_y}), we update the face-centered magnetic 
field according to
\begin{equation}\label{eq:stokes}\begin{array}{ccc}
\DS  B^{n+\HALF}_{x,i+\HALF} & = & \DS B^{n}_{x,i+\HALF}   
 - \frac{\Delta t^n}{2\Delta y_j}\Big(\Omega^z_{i+\HALF, j+\HALF} - \Omega^z_{i+\HALF, j-\HALF}\Big)\;,
          \\ \noalign{\medskip}
\DS  B^{n+\HALF}_{y,j+\HALF} & = & \DS B^{n}_{y,j+\HALF}   
 + \frac{\Delta t^n}{2\Delta x_i}\Big(\Omega^z_{i+\HALF, j+\HALF} - \Omega^z_{i-\HALF, j+\HALF}\Big)\;,
\end{array}
\end{equation}
and similarly after the corrector step.
The electromotive force $\Omega$ is collocated at cell corners and
is computed by straightforward arithmetic averaging:
\begin{equation}\label{eq:omega}
  \Omega^z_{i+\HALF, j+\HALF} = \frac{  \Omega^z_{i+\HALF,j}   + \Omega^z_{i, j+\HALF}
                                      + \Omega^z_{i+\HALF,j+1} + \Omega^z_{i+1,j+\HALF}}{4} \;,
\end{equation}
where, $\Omega^z_{i+\HALF,j} \equiv -f^{x,n}_{B_y,i+\HALF,j}$ and
$\Omega^z_{i,j+\HALF} \equiv f^{y,n}_{B_x,i,j+\HALF}$ are the $z$ components 
of the electric fields available at grid interfaces during the upwind step.
Despite its simplicity,  eq. (\ref{eq:omega}) lacks of directional bias 
and more sophisticated algorithms may be used to incorporate
upwind information in a consistent way, see \cite{LdZ04}, \cite{GS05}.    
For ease of implementation we will not discuss them here.

It is a straightforward exercise to verify that the 
$\nabla\cdot\A{B} = 0$ condition is preserved from one time step
to the next one, due to perfect cancellation of terms.
Notice also that, since $B_x$ is continuous at the $(i+\HALF,j)$ interface, only  
$\bar{B}_y$ and $\bar{B}_z$ need to be interpolated during the reconstruction 
procedure in the $x$-direction. A similar argument applies to $\bar{B}_x$
and $\bar{B}_z$ when interpolating along the $y$ coordinate. 

Since equation (\ref{eq:update}) evolves volume-averaged quantities, 
the zone-averaged magnetic field, $\bar{\A{B}}$, is computed at the 
beginning of the time step from the face-averaged magnetic fields 
using linear interpolation:
\begin{equation}\label{eq:bx_average}
 \bar{B}_{x}  = \frac{B_{x,i+\HALF} + B_{x,i-\HALF}}{2} \;,
\end{equation}
\begin{equation}\label{eq:by_average}
 \bar{B}_{y}  = \frac{B_{y,j+\HALF} + B_{y,j-\HALF}}{2} \;.
\end{equation}
Equations (\ref{eq:omega}), (\ref{eq:bx_average}) and 
(\ref{eq:by_average}) are second-order accurate in space.

\subsubsection{Summary}
%
%
We summarize our CTU constrained transport algorithm by the following 
steps:

\begin{enumerate}
\renewcommand{\theenumi}{(\arabic{enumi})}
\item At the beginning of the time step, form the volume averages 
      (\ref{eq:bx_average}) and (\ref{eq:by_average}) from 
      the face centered magnetic field.
\item Compute $x$ and $y$ limited slopes by interpolating 
      cell centered primitive variables according to eq. (\ref{eq:mc_lim})
      or (\ref{eq:vl_lim}).
\item\label{xpred} 
      Make a sweep along the $x$ direction. Form left and right states using 
      the first of eq. (\ref{eq:init_states}) with $B^{n}_{x,i+\HALF,L} =
      B^{n}_{x,i+\HALF,R}$ equal to the $x$ component of the face centered
      magnetic field;
 \begin{itemize}
   \item[-] use the Hancock step (\ref{eq:hancock_diff}) to compute 
         the $x$ derivative in eq. (\ref{eq:cent_pred_x}) and add the resulting 
         contribution to $U^{x,n+\HALF}$; 
   \item[-] compute the $\A{\cal L}^{x,n}$ Godunov operator by solving 
         Riemann problems at the $(i+\HALF,j)$ interfaces and add 
         the resulting contribution to $U^{y,n+\HALF}_{i,j}$.
 \end{itemize}
\item\label{ypred}
      Make a sweep along the $y$ direction. Form left and right states using 
      the second in eq. (\ref{eq:init_states}) with $B^{n}_{y,j+\HALF,L} 
      = B^{n}_{y,j+\HALF,R}$ equal to the $y$ component of the face centered 
      magnetic field;
 \begin{itemize}
   \item[-] obtain the $\A{\cal L}^{y,n}$ Godunov operator (\ref{eq:upwind_diff})
            by solving Riemann problems at the $(i,j+\HALF)$ interfaces; add 
            the resulting contribution to $U^{x,n+\HALF}_{i,j}$.
   \item[-] use the Hancock step relative to the $y$ direction to compute 
            the $y$ derivative and add it to $U^{y,n+\HALF}_{i,j}$; 
 \end{itemize}
\item Compute the time-centered area weighted magnetic field using  
      Stoke's theorem (\ref{eq:stokes}). 
      This concludes the predictor step.
\item Make a sweep along the $x$ direction with left and right 
      time-centered states given by the first equation in 
      (\ref{eq:pred_states}) with $B^{n+\HALF}_{x,i+\HALF,L} = 
      B^{n+\HALF}_{x,i+\HALF,R}$ 
      equal to the time centered face-averaged magnetic field computed via 
      Stoke's theorem.
      Obtain the $\A{\cal L}^{x,n+\HALF}$ Godunov operator. 
\item Repeat the previous step by sweeping along the $y$ direction. 
      Compute the $\A{\cal L}^{y,n+\HALF}$ Godunov operator. 
\item Update the cell-centered conservative variables using eq.
      (\ref{eq:update}) and the face-averaged magnetic field using 
      Stoke's theorem. 
\end{enumerate}

\subsection{One-dimensional test problems}\label{sec:1d}
%
%
%

One-dimensional problems are specifically designed to verify
the ability of the algorithm in reproducing the exact wave 
pattern. 
In what follows we present four shock-tube tests, already introduced 
by BA and dZBL, with left and right states given in Table \ref{tab:ic}. 
Computations are performed on the interval $[0,1]$ 
and the initial discontinuity is placed at $x = 0.5$. 
The final integration time is $t=0.4$.
Note that the constrained transport algorithm
is unnecessary, since eq. (\ref{eq:divB}) is trivially
satisfied in one-dimensional flows.
\begin{table}\begin{center}
\begin{tabular}{ccccccccc}
 Test &  $\rho$ & $p_g$  & $v_x$  & $v_y$ & $v_z$ & $B_x$ & $B_y$ & $B_z$ \\ \hline\hline 
1L    &   1     & 1      &  0     &  0    &  0    &  0.5  &  1    & 0     \\
1R    &   0.125 & 0.1    &  0     &  0    &  0    &  0.5  & -1    & 1     \\ \hline
2L    &   1     & 30     &  0     &  0    &  0    &  5    &  6    & 6     \\
2R    &   1     & 1      &  0     &  0    &  0    &  5    & 0.7   & 0.7   \\ \hline
3L    &   1     & $10^3$ &  0     &  0    &  0    &  10   &  7    & 7    \\
3R    &   1     & 0.1    &  0     &  0    &  0    &  10   & 0.7   & 0.7  \\ \hline
4L    &   1     & 0.1    &  0.999 &  0    &  0    &  10   &  7    &  7    \\
4R    &   1     & 0.1    & -0.999 &  0    &  0    &  10   & -7    & -7   \\ \hline
\end{tabular}
\caption{Initial conditions for the one-dimensional shock tube problems 
         presented in the text. In all test problems we adopt 
         a resolution of $1600$ uniform computational zone, covering 
         the interval $[0,1]$. Integration is carried until $t=0.4$.}
\label{tab:ic}
\end{center}\end{table}
  
\subsubsection{Problem 1}\label{sec:p1}
%
%

The first test problem, initially proposed by \cite{vP93}, 
is a relativistic extension of the \cite{BW88} magnetic shock tube. 
In analogy with the classical case we use the ideal equation 
of state (\ref{eq:eos}) with specific heat ratio $\Gamma = 2$.
The breakup of the initial discontinuity sets up a left-going fast
rarefaction wave, a left-going compound wave, a contact discontinuity,
a right-going slow shock and a right-going fast rarefaction wave.

\begin{figure}
 \includegraphics[width=80mm]{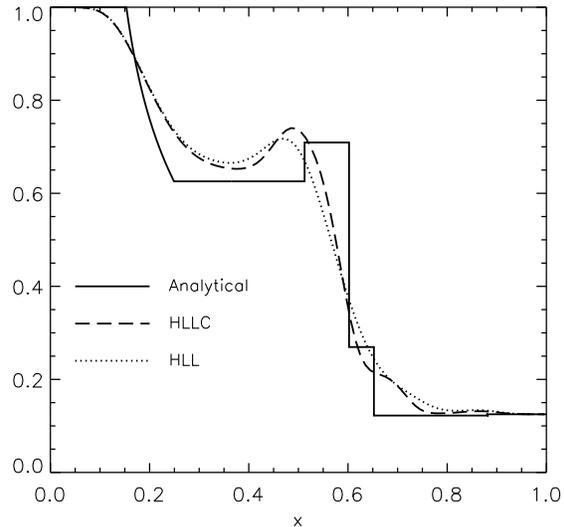}
 \caption{Comparison between the first-order HLL (dotted line) and the HLLC
         (dashed line) method for the first shock tube problem at $t=0.4$. 
          Only density profiles are shown. Computations were
          performed on $100$ computational zones with CFL = $0.8$.
          The solid line gives the analytic solution as computed by 
          \citet{GR05}.
          The major difference between the two approaches is the
          resolution of the contact wave.}
 \label{fig:flat1}
\end{figure}

\begin{figure}
 \includegraphics[width=85mm]{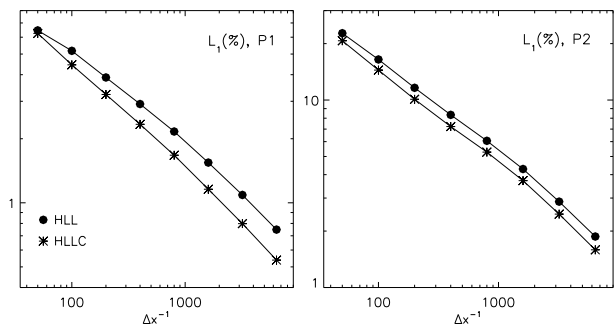}
 \caption{Discrete L1-norm density errors (in percent) computed 
          for the first-order scheme at different grid resolutions 
          using the HLLC (asterisks) and HLL (filled circles) solvers. 
          Computation have been performed for the first (left panel, P1) and 
          second (right panel, P2) problems on $50$, $100$, $200$, $400$,
          $800$, $1600$, $3200$ and $6400$ zones with $CFL = 0.8$.}
 \label{fig:res_1st}
\end{figure}

We compare, in Fig. \ref{fig:flat1}, the results 
obtained with the first-order HLL and HLLC solvers on $100$ uniform 
computational zones. The exact solution (given by the 
solid line) was obtained using the 
numerical code available from \cite{GR05}.  
The left going compound wave located at $x \approx 0.5$ 
is only visible in the numerical integration since 
the code used to generate the analytical solution 
(shown as the solid line in Fig. \ref{fig:flat1}) does 
not allow compound structures by construction.
As expected, the HLLC Riemann solver attains sharper representation of 
the contact discontinuity when compared to the HLL scheme. 
Because of the reduced smearing in proximity of the contact wave, 
neighboring structures such as the compound wave on the left and the 
slow shock on the right can be better resolved when using the HLLC
solver.  
Computations at different resolutions show, in fact, that the 
L-1 norm errors in density are reduced by roughly $20\div 30\%$
(see left panel in Fig. \ref{fig:res_1st}), with 
$L_1(\%)$ being, respectively, $0.53$ and $0.74$ 
for the HLLC and HLL solver at the highest resolution 
employed ($6400$ zones). 

\begin{figure*}
 \includegraphics[width=150mm]{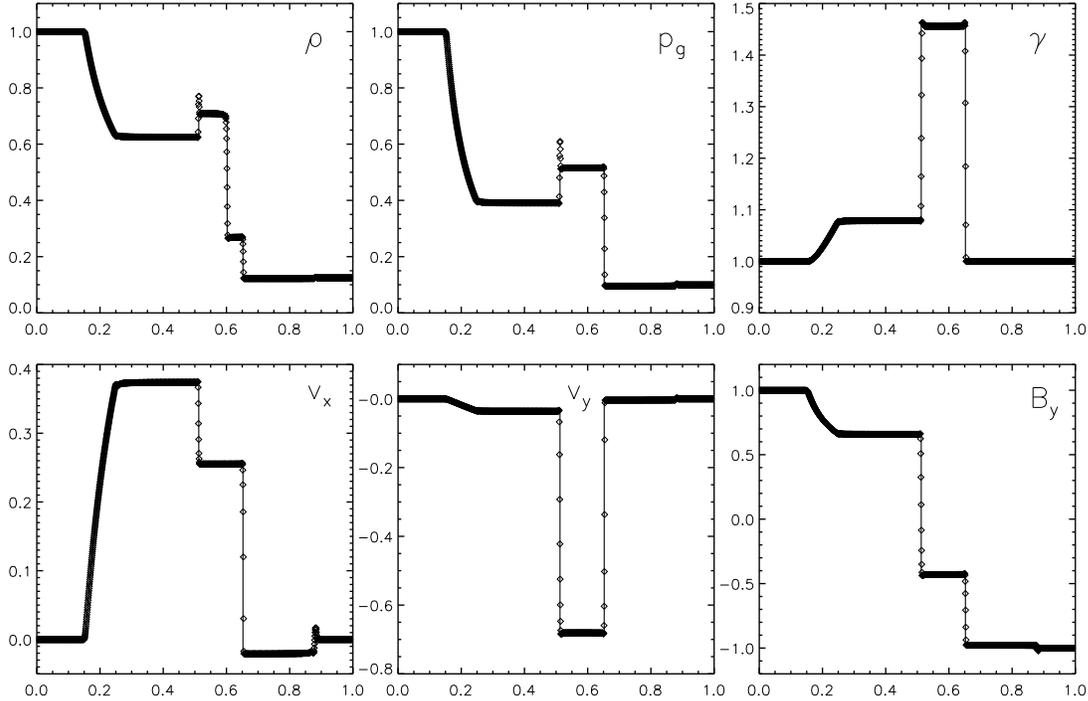}
 \caption{Relativisitc Brio-Wu shock tube problem. The second-order 
          scheme with the HLLC Riemann solver on $1600$ grid points
          and the MC limiter was used. 
          From left to right and top to bottom: proper density, 
          thermal pressure, Lorentz factor, normal and 
          transverse velocity components and transverse magnetic field.
          The Courant number is $0.8$.}
 \label{fig:sod1}
\end{figure*}
Fig. \ref{fig:sod1} shows the results obtained with the 
second-order scheme with the MC limiter, eq. (\ref{eq:mc_lim}), 
and the same Courant number, $CFL = 0.8$ on $1600$ grid points.
A direct comparison with the exact solution shows
that all discontinuities are correctly captured and resolved on 
few computational zones, owing also to the presence of a 
compressive limiter. 
In this respect, our second-order HLLC scheme provides similar 
results to those obtained with the third-order central ENO-HLL scheme 
by dZBL.  

The L-1 norm errors computed at different resolutions with 
the two different solvers differ by $\approx 10\div 20\%$,
see left panel in Fig. \ref{fig:res_2nd}.
When compared to the more sophisticated, characteristic-based
algorithm presented in BA, our results show slightly sharper
representation of the right-going slow shock and the contact 
discontinuity. 
Small overshoots appear in the Lorentz factor profile at the left
going compound wave and the right going slow shock.
More diffusive slope limiters do not exhibit this feature.

\subsubsection{Problem 2}\label{sec:p2}
%
%

The resulting wave pattern for this configuration is comprised of 
two left-going rarefaction fans (fast and slow) and two
right-going slow and fast shocks. 
The specific heat ratio used for this calculation is 
$\Gamma = 5/3$. 
The weak slow rarefaction located at $x\approx 0.53$ and the 
slow shock at $x\approx 0.86$ are separated 
by a contact discontinuity where the proper density
changes by a factor of $\sim 7$.
The velocity on either side of the contact wave is mildly 
relativistic, with a maximum Lorentz factor of $\approx 1.36$.

\begin{figure}
 \includegraphics[width=85mm]{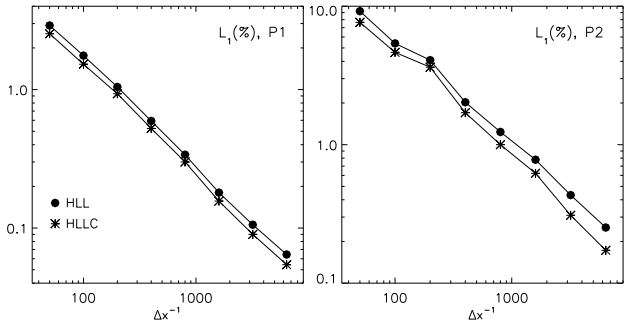}
 \caption{Discrete L1-norm error ($10^{-2}$) for density 
          computed for the second order scheme at different
          resolutions, see Fig. \ref{fig:res_1st}.}
 \label{fig:res_2nd}
\end{figure}
\begin{figure}
 \includegraphics[width=80mm]{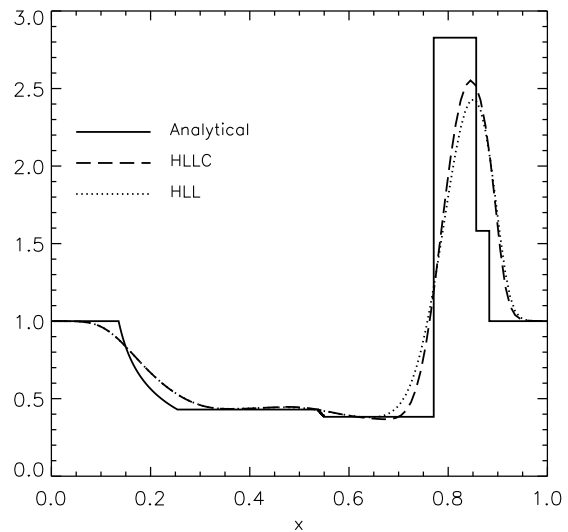}
 \caption{Comparison between the first-order HLL (dotted line) and the HLLC
         (dashed line) method for the second shock tube  at $t=0.4$. 
          Density profiles are shown. Computations were
          performed on $100$ computational zones with CFL = $0.8$.
          The solid line gives the analytic solution as computed by 
          \citet{GR05}.}
 \label{fig:flat2}
\end{figure}

The improvement offered by the HLLC Riemann solver over the HLL approach
in the resolution of the contact wave is evident from Fig. \ref{fig:flat2}, 
where we compare the density profiles obtained with the first order 
schemes against the analytical solution. 

\begin{figure*}
 \includegraphics[width=150mm]{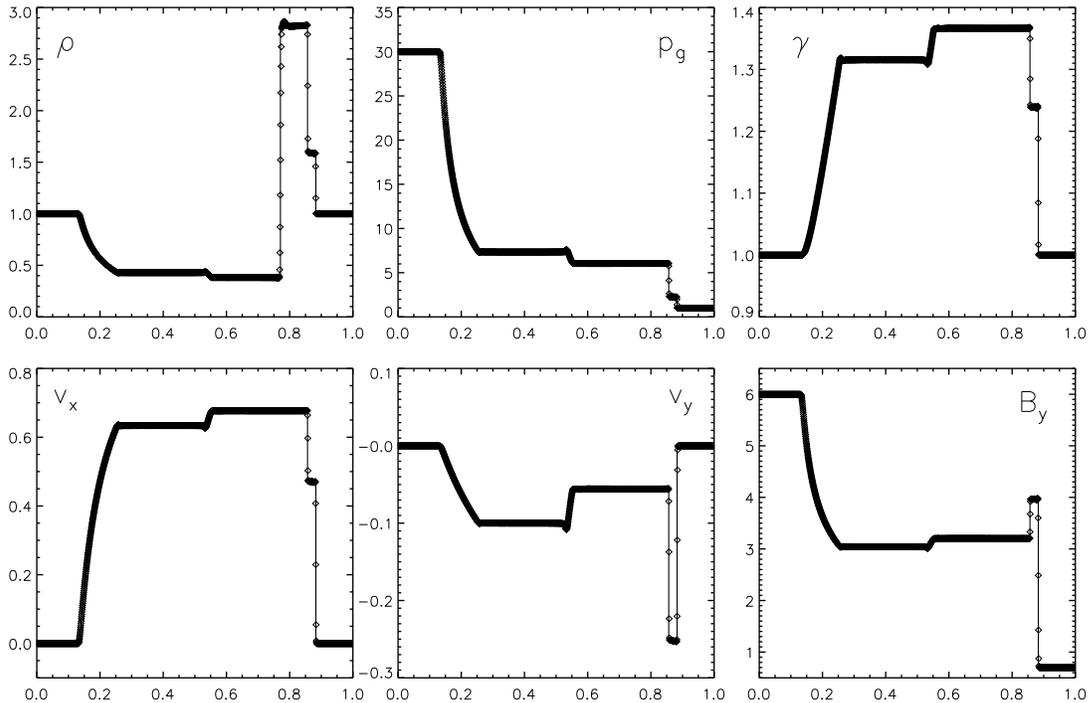}
 \caption{Solution of the mildly relativistic blast wave problem (test 2)
          computed with the second-order HLLC scheme and the MC limiter. 
          A Courant number of $0.8$ and $1600$ grid zones were used in the computation.}
 \label{fig:sod2}
\end{figure*}

Computations obtained with the second-order limiter (\ref{eq:mc_lim}) 
show excellent agreements with the analytical profiles, see Fig. 
\ref{fig:sod2}.
Our single-step HLLC scheme attain considerably sharper 
resolution than the results obtained by previous calculations.
The two right-going shocks, for instance, are smeared over  
$\sim 3$ grid points, approximately half of the resolution shown 
in BA and dZBL.
Moreover, the smearing of the contact wave is considerably reduced 
when compared to the HLL scheme in dZBL ($\sim 10$ zones vs.
$\sim 14$). Similar overshoots, though, appear at the right of 
contact mode.  

The discrete L-1 errors for different grid sizes are shown 
in the right panel of Fig. \ref{fig:res_2nd}, where, at the 
maximum resolution employed ($6400$ zones) the HLLC and HLL errors 
reduce to  $0.17 \%$ and $0.25\%$, respectively.

\subsubsection{Problem 3}\label{sec:p3}
%
%

\begin{figure*}
 \includegraphics[width=150mm]{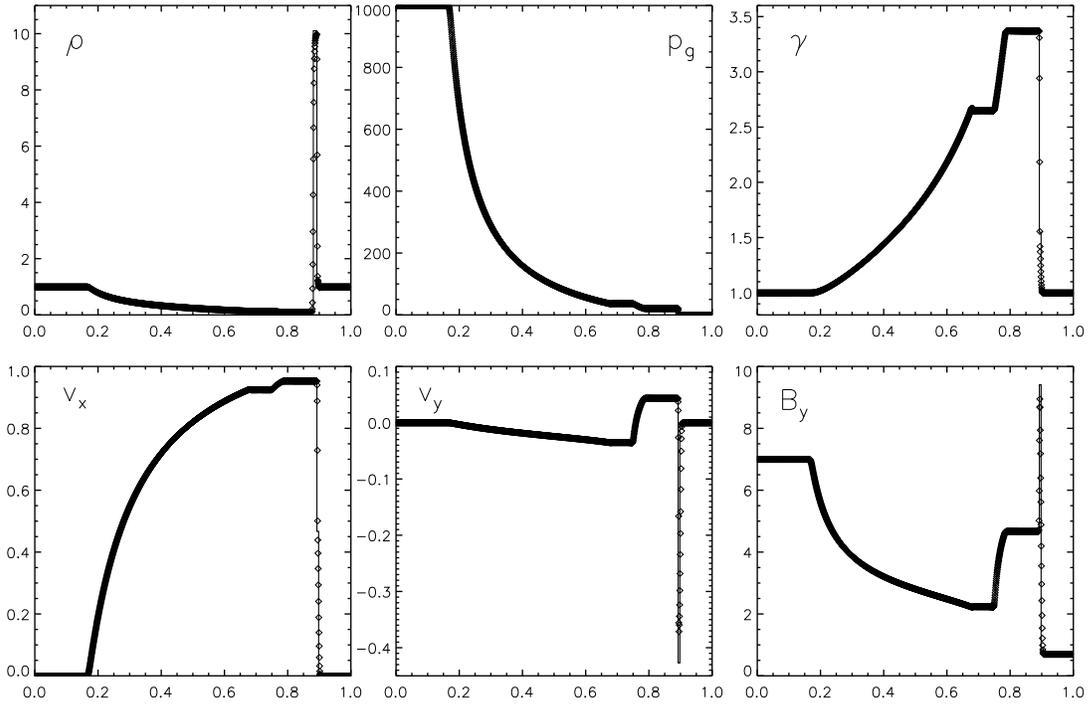}
 \caption{Strong blast wave problem (test 3) on $1600$ grid points.
          A Courant number of $0.8$ and the Van Leer limiter were used.}
 \label{fig:sod3}
\end{figure*}

The configuration for this test is similar to the previous problem,
but a higher pressure jump separates the initial left and right states,
see Table \ref{tab:ic}. Only the second-order scheme with the 
Van Leer limiter (\ref{eq:vl_lim}) and a Courant number of $0.8$ 
has been employed. The ideal equation of state (\ref{eq:eos}) 
with $\Gamma = 5/3$ is used.
The ensuing wave pattern shows a stronger relativistic configuration, 
with a maximum Lorentz factor of $\sim 3.37$, see Fig. \ref{fig:sod3}.
The presence of magnetic fields makes the problem even more
challenging than its hydrodynamical counterpart (see test 3 in
paper I), since the contact wave, slow and fast shocks now propagate 
extremely close to each other.
As a result, a thin density shell sets up between the contact mode
and the slow shock. The higher compression factor (more than $100$) 
follows from a more pronounced relativistic length 
contraction effect. 
At the resolution of $1600$ grid zones, the relative 
error in the density peak ($\rho_{\max} \approx 9.98$) is $1.2 \%$.
A second thin shell-like structure forms between the 
slow and fast shocks, as can be seen in the profiles in 
Fig. \ref{fig:sod3}.
The peaks achieved in the transverse components of velocity 
($\approx -0.37$) and magnetic field ($\approx 8.95$) 
achieve, respectively, $87\%$ and $95\%$ of their exact values.
The small shell thickness, however, still prevents 
a clear resolution of the two right going shocks, 
visible in the exact solution.
This demonstrates that relativistic magnetized flows can develop
rich and complex features difficult to resolve on a grid of fixed
size.
Similar conclusions have been drawn by previous investigators. 

Results obtained with the HLL solver (not shown here) indicates that
the resolution attained at the contact discontinuity is 
equivalent.
Therefore, as it was also pointed out in paper I, we conclude that, 
for strong blast waves where relativistic contraction effects produce 
closely moving discontinuities, the HLL and HLLC schemes produce nearly 
identical results.

\subsubsection{Problem 4}\label{sec:p4}
%
%

\begin{figure*}
 \includegraphics[width=150mm]{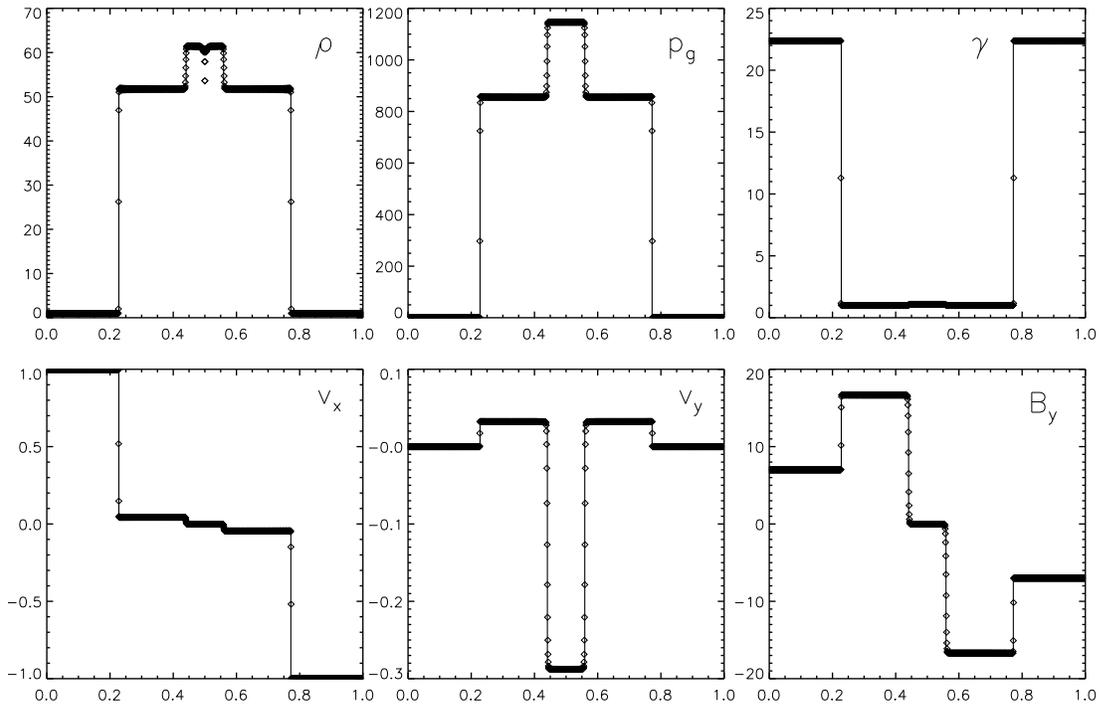}
 \caption{Relativistic shock reflection problem at $t = 0.4$ on $1600$ 
          computational cells. The initial Lorentz factor is 
          $\gamma \approx 22.4$. Integration has been carried with 
          the Van Leer limiter (except near strong shocks where the minmod
          limiter was used) and a Courant number of $0.8$.
          Notice the wall heating 
          problem, evident in the density profile.}
 \label{fig:sod4}
\end{figure*}
The collision of two relativistic streams is considered in 
the fourth test problem.
The initial impact produces two strong relativistic fast shocks
propagating symmetrically in opposite direction about the impact 
point, $x=0.5$, see Fig. \ref{fig:sod4}. 
Two slow shocks delimiting a high pressure region in the center 
follow behind.

Computations are carried out with $CFL = 0.8$ and the Van Leer
limiter, eq. (\ref{eq:vl_lim}). 
Spurious oscillations in vicinity of strong shocks are reduced 
by switching to the more diffusive minmod limiter, see
\S\ref{sec:shockflattening}.
No contact waves are present in the problem and, not
surprisingly, the quality of our solution is essentially
the same obtained by previous authors: the fast shocks
are resolved in $2\div 3$ cells, whereas the slow
shocks are smeared out over $5\div 6$ zones. Very similar 
patterns are observed in the work of BA and dZBL.

It is well known that Godunov-type schemes suffer from 
a common pathology, often found in these type of problems.
In the classical case, this has been recognized
for the first time by \cite{Noh87}.
The wall heating problem, in fact, consists in an undesired 
entropy buildup in a few zones around the point of symmetry.
Our scheme is obviously no exception as it can be 
inferred by inspecting the density profile 
in Fig. \ref{fig:sod4}.

We repeated the test with the HLL scheme and 
found that this pathology is worse when the HLLC scheme 
is used. The relative numerical undershoot in density,
in fact, were found to be $\sim 5\%$ for the HLL and 
$\sim 12\%$ for the HLLC scheme. Since similar
errors were also reported by BA, and the same conclusions 
have been drawn in paper I, we raise the question as
to whether the degree of this pathology grows with the 
complexity of the Riemann solver.
Future, more specific works should address this problem.

\subsection{Two-dimensional test problems}\label{sec:multid}
%
%
%

Multi-dimensional numerical computations of magnetized flows 
are notoriously more challenging, due to the necessity to preserve the 
divergence-free constraint (\ref{eq:divB}). 
In what follows, we consider three test problems: a cylindrical blast 
wave test, the interaction of a strong magnetosonic shock with a cloud
and the propagation of an axisymmetric jet in cylindrical coordinates. 
 
\subsubsection{Cylindrical Blast Wave}\label{sec:p5}
%
%

\begin{figure*}
 \includegraphics[width=150mm]{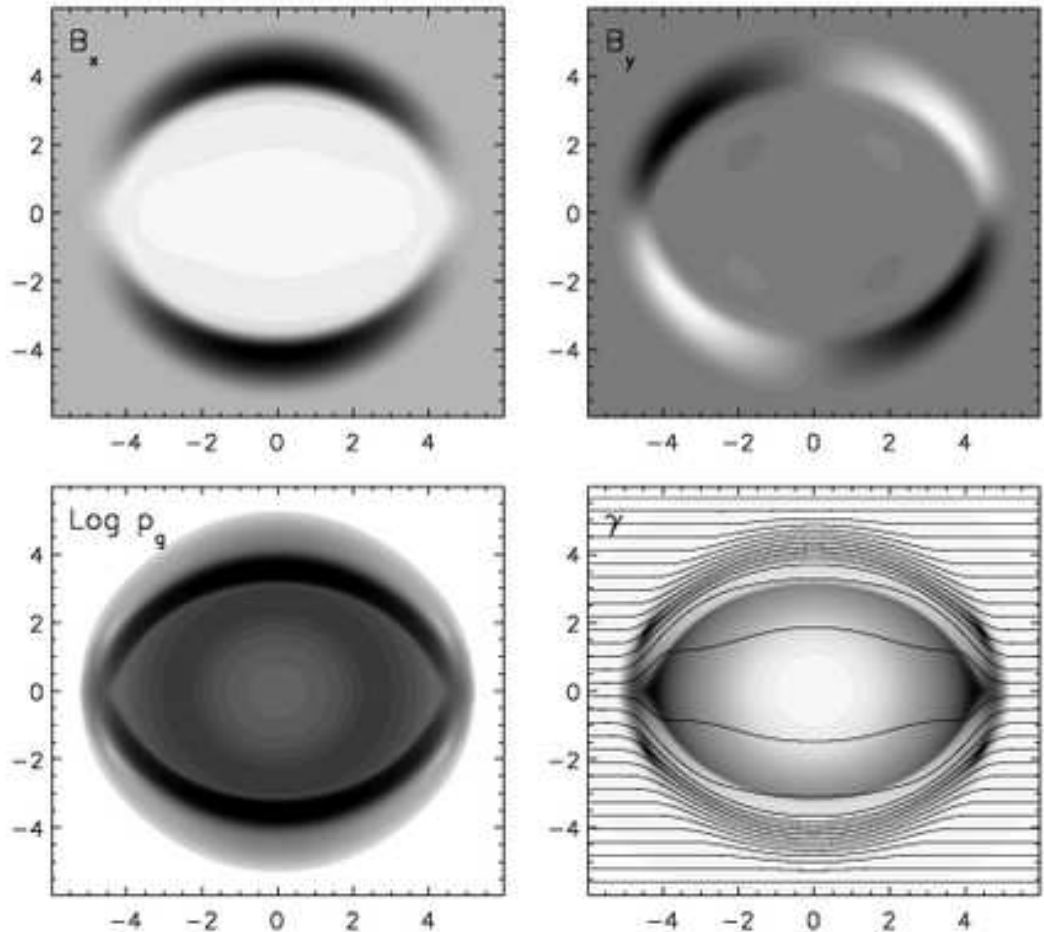}
 \caption{Gray scale levels of the $x$ component of magnetic field 
          (top left), $y$ component of magnetic field (top right), 
          gas pressure logarithm (bottom left) and 
          Lorentz factor (bottom right) for the cylindrical 
          blast wave with relatively weak magnetic field at $t=4$. 
          Magnetic field lines are plotted on top of the Lorentz
          factor distribution.
          Following KO, we use 32 equally spaced contour levels between
          $0.008$ and $0.35$ (for $B_x$), $-0.18$ and $0.18$ (for
          $B_y$), $-4.5$ and $-1.5$ (for $\textrm{Log}\;p_g$), 
          $1$ and $4.57$ (for $\gamma$). 
          }
 \label{fig:blast_lo}
\end{figure*}
\begin{figure*}
 \includegraphics[width=150mm]{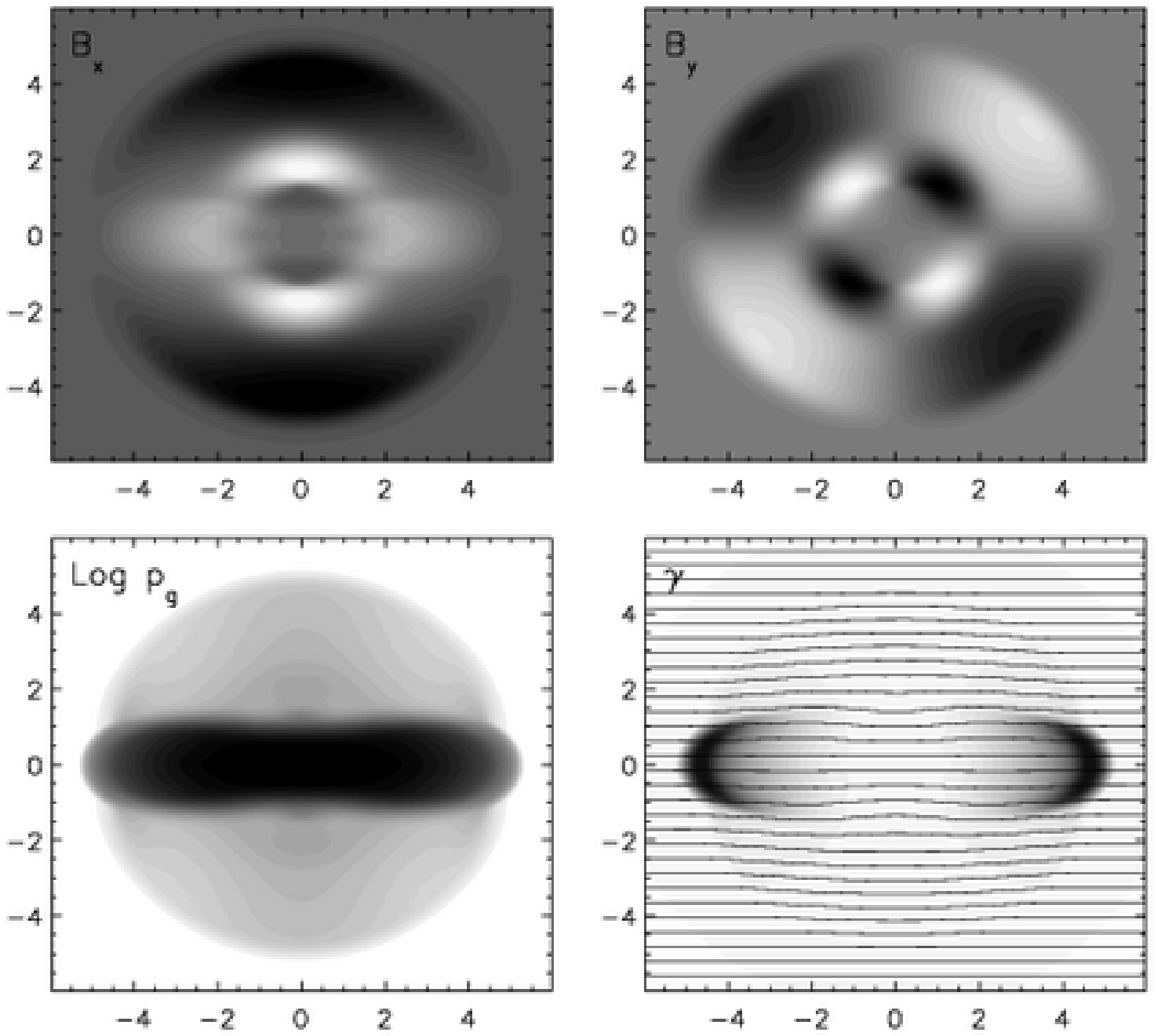}
 \caption{Cylindrical explosion for the strong magnetic
          field case ($B_x = 1$).
          We use 32 equally space contour levels between 
          $0.793$ and $1.116$ (for $B_x$), $-0.09$ and $0.09$ (for $B_y$),
          $-4.52$ and $-1.02$ (for $\textrm{Log}\;p_g$), 
          $1$ and $4.23$ (for $\gamma$).}
 \label{fig:blast_hi}
\end{figure*}
Cylindrical explosions in cartesian coordinates are
particular useful in checking the robustness of the code and the 
algorithm response to different kinds of degeneracies. 
Here we follow the same setup adopted by KO, where the 
square $[-6, 6]\times[-6, 6]$ is filled with a uniform 
($\rho = 10^{-4}$, $p_g=3\cdot10^{-5}$), initially static ($\A{v} = \A{0}$) 
medium, threaded by a constant magnetic field $\A{B} = (B_x,0)$. 
The circular region $\sqrt{x^2 + y^2} < 0.08$ is initialized 
with constant higher density and pressure values, 
$\rho = 0.01$ and $p_g = 1$ decreasing linearly
for $0.08\le r \le 1$. 
We adopt the ideal equation of state (\ref{eq:eos}) with specific heat 
ratio $\Gamma = 4/3$.
We consider two setups, corresponding to a relatively weak 
magnetic field $B_x = 0.1$ and a strong field $B_x = 1$.
Figures \ref{fig:blast_lo} and \ref{fig:blast_hi} 
show the magnetic field distribution, thermal
pressure and Lorentz factor for the two configurations at $t = 4$. 
Computations are carried using the van Leer limiter, eq. (\ref{eq:vl_lim}),
together with the multidimensional limiting procedure described in 
\S\ref{sec:mdlimit} on $200\times200$ uniform grid zones.
The Courant number is $0.4$.

The expanding region is delimited by a fast forward shock propagating
(nearly) radially at almost the speed of light.
In the weak field case, a reverse shock delimits the inner region 
where expansion takes place radially. 
Magnetic field lines are squeezed in the $y$ direction  
building up a shell of higher magnetic pressure. 
In the $x$ direction the motion of the gas is not 
hindered by the presence of the field and it achieves a higher 
Lorentz factor ($\gamma_{\max} = 4.39$).      
In the strong field case, the expansion is magnetically confined
along the $x$ direction and the outer fast shock has reduced
amplitude. The maximum Lorentz factor is $\gamma_{\max} = 4.02$.

We point out that numerical integrations for this test were 
possible only by locally redefining the total energy 
at the end of the time step:
\begin{equation}\label{eq:new_E}
 E \rightarrow E + \frac{\bar{\A{B}}_{fa}^2 - \bar{\A{B}}_c^2}{2}\;,
\end{equation}
where $\bar{\A{B}}_c$ is the cell-centered magnetic field obtained after 
the Godunov step, whereas $\bar{\A{B}}_{fa}$ 
is the new magnetic field obtained by averaging the face centered
values given by (\ref{eq:stokes}). 
Notice that equation (\ref{eq:new_E}) only redefines the energy contribution
of the magnetic field that is not directly coupled to the velocity, 
see eq. (\ref{eq:cons_var_E}) and thus may be regarded as a first-order
correction.  
In this respect, the energy correction
we propose is the same usually adopted in CT schemes, see \cite{BS99},
\cite{Toth00}.
Although this optional step results in a slight loss of energy conservation
at the discretization level, it was nevertheless found to become 
particularly useful in problems where the magnetic pressure dominates over
the thermal pressure by more than two order of magnitudes. 

\subsubsection{Relativistic Shock-Cloud Interaction}\label{sec:p6}
%
%

The interaction of a strong relativistic fast shock with a  
cloud is considered on the unit square $[0,1]\times[0,1]$
in 2-D cartesian coordinates $(x,y)$.
This problem has been extensively used for testing classical MHD codes
see \cite[][and references therein]{DW94,Toth00}.
Here we consider a relativistic extension adopting a somewhat 
different initial condition, with magnetic field 
orthogonal to the slab plane. 
The shock wave travels in the positive $x$-direction and is initially 
located at $x=0.6$. Upstream, for $x > 0.6$, the flow is highly 
supersonic with pre-shock values given by $(\rho, \gamma_x, p_g, B_z)_\textrm{pre} = 
(1, 10, 10^{-3}, 0.5)$, where $\gamma_x = (1 - v_x^2)^{-\HALF}$.
In this reference frame, shocked material is at rest with values given by 
\begin{equation}
 \left(\begin{array}{c}
  \rho  \\ \noalign{\medskip}
  p_g   \\ \noalign{\medskip}
  B_z   
 \end{array}\right)_{\textrm{post}} = 
 \left(\begin{array}{c}
  42.5942      \\ \noalign{\medskip}
  127.9483     \\ \noalign{\medskip}
 -2.12971     \\ \noalign{\medskip}
 \end{array}\right) \;.
\end{equation}
Notice that the magnetic field carries a rotational discontinuity
and the compression factor of density across the shock in not 
limited to $7$ (we use $\Gamma = 4/3$) as in the classical case, 
but achieves a much higher value ($\approx 43$). 
This feature is unique to relativistic flows.  

\begin{figure}
 \includegraphics[width=80mm]{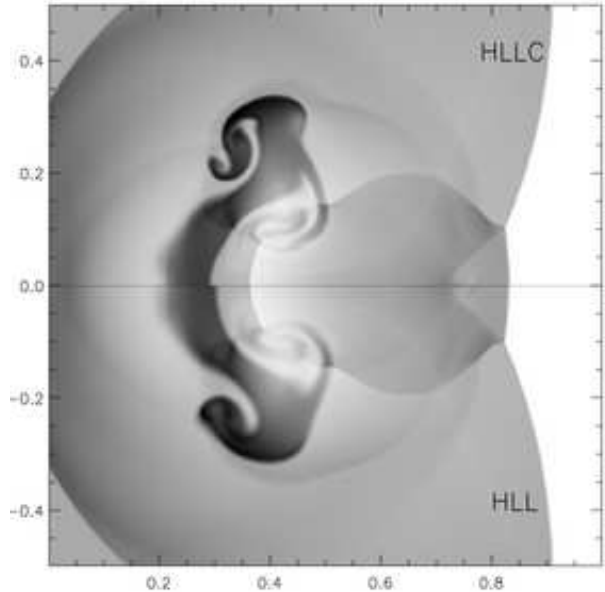}
 \caption{Density gray scale map of the interaction between a 
          strong shock and a cloud at $t=1$. 
          The upper and lower halves show the solutions computed 
          with HLLC and HLL solvers, respectively, on $400\times200$
          zones, with CFL = $0.4$ and the MC limiter. 
          Shock-flattening has been used to prevent
          spurious oscillations in proximity of the slow moving shock.}
 \label{fig:sc}
\end{figure}
A circular density clump with $\rho = 10$ and radius $r = 0.15$ 
is placed ahead of the shock front, centered at $(x,y) = (0.8,0.5)$.
Transverse velocities $v_y$ and $v_z$ and the $x$ and $y$ components
of magnetic field are set to zero everywhere.
We use $400\times 200$ computational zones, by assuming reflecting 
boundary at $y = 0.5$ and free flow across the remaining boundaries.
The MC limiter, eq. (\ref{eq:mc_lim}), is employed everywhere
except in proximity of strong shocks where we 
revert to the minmod limiter, see \S\ref{sec:shockflattening}.
The Courant number is $0.4$.

Shortly after the impact, the cloud undergoes strong 
compression with the density rising by a factor of more than $20$.
The collision generates a bow fast shock propagating in the shocked
material and a reverse shock is transmitted into the cloud.
After the transmitted shock reaches the back of the cloud, the
two bent parts of the original incident shock join back together 
and complicated wave pattern emerges.  
By $t=1$ the cloud is completely wrapped by the incident shock, 
and the cloud expands in the form of a mushroom-shaped shell, 
see upper half of Fig. \ref{fig:sc}.
The solution computed with the HLL solver (lower half in 
Fig. \ref{fig:sc}) show similar structures, although 
the amount of numerical viscosity is considerably higher.

Notice that, because of the assumed slab symmetry, the condition 
$\A{v}\cdot\A{B} = 0$ is preserved in time
and the solution to the Riemann problem at each interface 
consists of a three wave pattern: 
two fast waves separated by a tangential discontinuity.
In this regard, our HLLC solver provides a better approximation 
of the full wave structure.

\subsubsection{Relativistic Jet}\label{sec:p8}
%
%

As a final example, we consider the propagation of an axisymmetric 
jet in cylindrical coordinates $(r,z)$.
The configuration adopted here corresponds to model C2-pol-1 in \cite{LAAM05}. 

The domain $[0,12]\times[0,50]$ (in units of jet beam) is initially 
filled with a static uniform distributions of density, gas pressure and 
magnetic field, given respectively by
\begin{equation}
 \rho_a = 1 \,,\quad 
 p_a = \frac{\eta v_b^2}{\Gamma(\Gamma-1)M^2 - \Gamma v_b^2}
 \, ,\quad B_z = \sqrt{2p_a}.
\end{equation}
The numerical value of $p_a$ follows from the definitions of 
the beam Mach number $M = v_b/c_s = 6$, jet to ambient density 
ratio $\eta = 10^{-2}$ and beam axial velocity $v_b=0.99$.
The ideal equation of state (\ref{eq:eos}) is used with $\Gamma = 5/3$.
The jet nozzle is located at the lower boundary $r \le 1$, $z=0$, 
where boundary conditions are held constant in time, 
$(\rho, v_r, v_z, B_r, B_z, p_g) = (\eta, 0, v_b, 0, B_z, p_a)$.
For $r>1$ we prescribe boundary values with antisymmetric
profiles for axial velocity and radial magnetic field.
Symmetric profiles are imposed on the remaining quantities.
This configuration corresponds to a twin counter jet propagating
in the opposite direction.
Outflow boundaries are imposed on all other sides, except 
at $r=0$ where reflecting boundary conditions are used.
We employ a uniform resolution of $20$ zones per beam radius 
and carry integration until $t = 126$ with $CFL = 0.4$.

\begin{figure}
 \includegraphics[width=85mm]{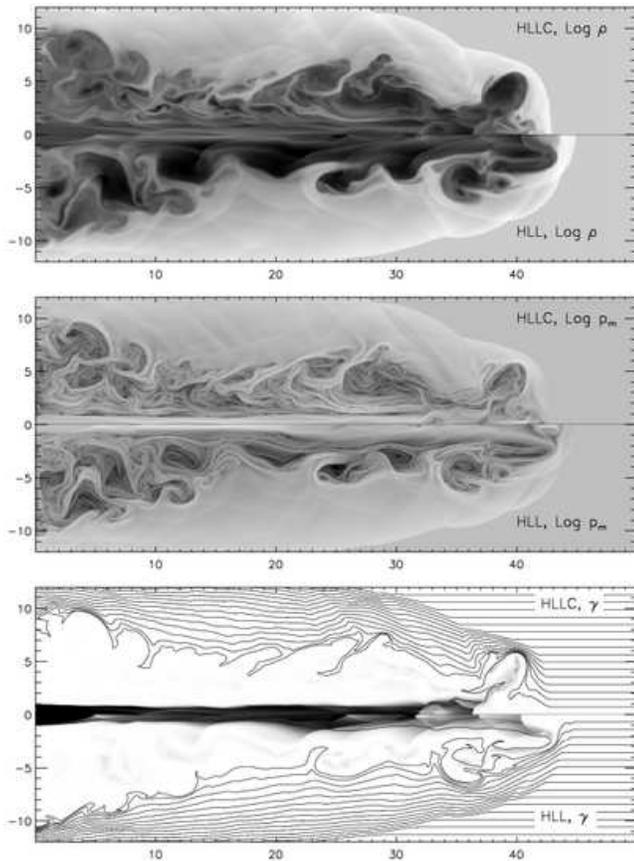}
 \caption{Gray scale images of density (top panel), magnetic 
          pressure (middle panel) and Lorentz factor (bottom panel) for
          the axisymmetric jet. The upper (lower) half in each panel
          refers to the integration carried with the HLLC (HLL) solver.
          Both integrations were carried till $t = 126$ with CFL = $0.8$
          and the Van Leer limiter.
          An ideal equation of state is used with $\Gamma = 5/3$.
          Magnetic field lines are plotted on top of the Lorentz 
          factor gray-scale images.}
 \label{fig:jet}
\end{figure}
The results are shown in Fig. \ref{fig:jet}, where we display 
density logarithm (upper panel), magnetic pressure (middle panel) and 
Lorentz factor distributions (lower panel).
In each panel, the upper and lower halves show the solutions obtained
with the HLLC and HLL solvers, respectively. 
As we already pointed out in the non magnetic case (Paper I), the 
HLLC integration features considerably less amount numerical 
diffusion as evident from the richness in small scale structures,
notably in the density distribution. 
In fact, density is the physical quantity more sensitive to the 
introduction of the tangential wave in the Riemann 
solver. Comparing our results with those of \cite[][see their Fig. 5]{LAAM05}
we can observe that our solution has a similar 
(or even larger) richness in fine structure details at half the resolution 
(20 ppb in our case, 40 ppb in their case).

\section{Conclusions}
%
%
%
%

An HLLC approximate Riemann solver has been developed 
for the relativistic magnetohydrodynamic equations.
The new approach improves over the single state HLL solver 
in the ability to capture exactly isolated tangential and 
contact discontinuities. 
Several test problems in one and two dimensions demonstrate
better resolution properties and a reduced amount of the
numerical diffusion inherent to the averaging process of the 
single state HLL scheme.
The solver is well-behaved for strictly two-dimensional flows,
although applications to genuinely three-dimensional problems 
may suffer from a pathological singularity when the 
component of magnetic field normal to a zone interface
approaches zero. This feature does not persist in the classical
limit. 

Multidimensional integration has been formulated in a 
versatile and efficient way within the framework 
of the corner transport upwind (CTU) method.
The algorithm is stable up to Courant numbers of $1$ and 
preserves the divergence-free condition 
via constrained transport evolution of the magnetic field.
The additional computational cost and the numerical 
implementation in an existing relativistic MHD code are 
minimal.

%
%

\appendix
%
%

\section{}

\subsection{Shock Flattening}\label{sec:shockflattening}
%
%

For strong shocks, we found that the  
one-dimensional prescriptions (\ref{eq:mc_lim}) or (\ref{eq:vl_lim})
can still produce spurious numerical oscillations eventually 
leading to the occurrence of negative pressures.
A weak form of flattening is introduced by replacing eq. 
(\ref{eq:mc_lim}) or (\ref{eq:vl_lim}) with the minmod limiter 
whenever a strong shock is detected. 
In order for the latter condition to hold, we require that
both $\nabla\cdot\A{v} < 0$ and $\chi_{\min} = 0$,  
where $\nabla\cdot\A{v}$ is computed by central differences whereas
\begin{equation}
 \chi_{\min} = \min\left(\chi^x_{i+1,j},\chi^x_{i,j},\chi^x_{i-1,j},
                         \chi^y_{i,j+1},\chi^y_{i,j},\chi^y_{i,j-1}\right) \,.
\end{equation}
The switches $\chi^x$ and $\chi^y$ are designed as follows
\begin{equation}
 \chi^x_{i,j} = \left\{\begin{array}{cc}
  1 & \DS \; \textrm{if}\quad \frac{p_{i+1,j} - p_{i-1,j}}
                               {\min\left(p_{i+1,j}, p_{i-1,j}\right)} \le \epsilon \;,\\ \noalign{\medskip}
  0 & \DS \; \textrm{otherwise} \;,
\end{array}\right.
\end{equation}
\begin{equation}
 \chi^y_{i,j} = \left\{\begin{array}{cc}
  1 & \DS \; \textrm{if}\quad \frac{p_{i,j+1} - p_{i,j-1}}
                               {\min\left(p_{i,j+1}, p_{i,j-1}\right)} \le \epsilon \;,\\ \noalign{\medskip}
  0 & \DS \; \textrm{otherwise} \;,
\end{array}\right.
\end{equation}
where we set $\epsilon = 5$ in all computations presented 
in this paper.

\subsection{Multidimensional Limiting}
\label{sec:mdlimit}
%
%

Occasionally, we found that strong shocks propagating 
obliquely to the grid in highly magnetized media may 
benefit from an additional form of limiting, based 
on genuinely multidimensional constraints.
When needed, we enforce the maximum and minimum interpolated values 
in each cell $(i,j)$ to lie within the bounds provided by the four 
neighboring zones $(i+1,j), (i-1,j), (i,j+1), (i,j-1)$.
Specifically, denote with $\hat{q}^{\max}$ and 
$\hat{q}^{\min}$ the maximum and minimum values of 
$q\in\A{V}$ in these cells.
Once the limited slopes $\delta_x q$ and $\delta_y q$
have been computed according to (\ref{eq:mc_lim}) or
(\ref{eq:vl_lim}), we apply the correction  
\begin{equation}
 \delta_xq \rightarrow \tau\delta_xq\;, \quad
 \delta_yq \rightarrow \tau\delta_yq\;, \quad
\end{equation}  
where the multi-dimensional limiter $\tau$ is constructed 
as in \cite{Balsara04}:
\begin{equation}
 \tau = \min\left(1,\psi\min\left(
       \frac{\hat{q}^{\max} - q}{\delta^{\max}},
       \frac{q - \hat{q}^{\min}}{\delta^{\min}}
     \right)\right)   \;,
\end{equation}
with $\delta^{\max} = \max(|\delta_x q|,|\delta_y q|)$, 
$\delta^{\min} = \min(|\delta_x q|,|\delta_y q|)$.
We set $\psi = 2$ for density and magnetic field, $\psi = 3/4$ for
velocity and $\psi = 1$ for thermal pressure.

\label{lastpage}

\begin{thebibliography}{}

\bibitem[Aloy et al.(2000)]{Aloy_etal00} 
  Aloy, M.~A., M{\"u}ller, E., Ib{\'a}{\~n}ez, J.~M., Mart{\'{\i}}, J.~M., \& MacFadyen, A.\ 
  2000, ApjL, 531, L119 

\bibitem[Aloy et al.(2002)]{Aloy_etal02}
   Aloy, M.-A., Ib{\'a}{\~n}ez, J.-M., Miralles, J.-A., \& Urpin, V.\ 
   2002, Astronomy and Astrophysics, 396, 693 

\bibitem[Anile \& Pennisi (1987)]{AP87}
  Anile, M., \& Pennisi, S.   \
  1987, Ann. Inst. Henri Poincar{\'{e}}, 46, 127

\bibitem[Anile (1989)]{Anile89}
  Anile, A.~M.\ 
  1989, Relativistic Fluids and Magneto-fluids (Cambridge: Cambridge University Press), 55

\bibitem[Balsara \& Spicer (1999)]{BS99}
  Balsara, D.~S., Spicer, S. D.\ 
  1999, J. Comput. Phys, 149, 270

\bibitem[Balsara(2001)]{Balsara01}
  Balsara, D.~S.\ 
  2001, ApJS, 132, 83 (BA)

\bibitem[Balsara(2004)]{Balsara04}
  Balsara, D.~S.\ 
  2004, ApJS, 151, 149

\bibitem[Brio \& Wu (1988)]{BW88}
  Brio, M., \& WU, C.-C. \
  1988, J. Comput. Phys., 75, 400

\bibitem[Bruenn(1985)]{Bruenn85} 
  Bruenn, S.~W.\ 
  1985, ApJS, 58, 771

\bibitem[Bucciantini et al.(2005)]{BZAV05} 
  Bucciantini, N., del Zanna, L., Amato, E., \& Volpi, D.\ 
  2005, Astronomy \& Astrophysics, 443, 519  

\bibitem[Colella(1990)]{Colella90}
  Colella, P.
  1990, J. Comput. Phys., 87, 171

\bibitem[Courant et al.(1928)]{CFL28}
  Courant, R., Friedrichs, K.~O. \& Lewy, H.\
  1928,  Math. Ann., 100, 32

\bibitem[Dai \& Woodward (1994)]{DW94}
  Dai, W., \& Woodward, P.R. \
  1994, ApJ, 436,776

\bibitem[Davis(1988)]{Davis88}
  Davis, S.F.\ 
  1988, SIAM J. Sci. Statist. Comput., 9, 445

\bibitem[Del Zanna et al.(2003)]{dZBL03}
  Del Zanna, L., Bucciantini, N., \& Londrillo, P.\ 
  2003, Astronomy \& Astrophysics, 400, 397, (dZBL) 

\bibitem[Dimmelmeier et al.(2002)]{DFM02} 
  Dimmelmeier, H., Font, J.~A., M{\" u}ller, E.\ 
  2002, Astronomy and Astrophysics, 393, 523 

\bibitem[Duez et al.(2005)]{DLSS05} 
  Duez, M.~D., Liu, Y.~T., Shapiro, S.~L., \& Stephens, B.~C.\ 
  2005, Phys. Rev. D, 72, 024028 

\bibitem[Einfeldt et al.(1991)]{EMRS91}
  Einfeldt, B., Munz, C.D., Roe, P.L., and Sj{\" o}green, B.\ 
  1991, J. Comput. Phys., 92, 273

\bibitem[Elvis et al.(2002)]{ERZ02}
  Elvis, M., Risaliti, G., \& Zamorani, G.\ 
  2002, ApJL, 565, L75 

\bibitem[Evans \& Hawley(1988)]{EH88} 
  Evans, C.~R., \& Hawley, J.~F.\ 
  1988, apj, 332, 659 
 
\bibitem[Gammie et al.(2003)]{GmKT03} 
  Gammie, C.~F., McKinney, J.~C., \& T{\'o}th, G.\ 
  2003, ApJ, 589, 444 

\bibitem[Gardiner \& Stone(2005)]{GS05} 
  Gardiner, T.A., \& Stone, J.M.\ 
  2005, Journal of Computational Physics, 205, 509 

\bibitem[Giacomazzo \& Rezzolla(2005)]{GR05}
  Giacomazzo, B., \& Rezzolla, L.\
  2005, J. Fluid Mech., xxx

\bibitem[Gurski(2004)]{Gurski04}
  Gurski, K.F. \
  2004, SIAM J. Sci. Comput, 25, 2165

\bibitem[Harten et al.(1983)]{HLL83}
  Harten, A., Lax, P.D., and van Leer, B.\ 
  1983, SIAM Review, 25(1):35,61

\bibitem[Koldoba et al.(2002)]{KKU02}
  Koldoba, A.~V., Kuznetsov, O.~A., \& Ustyugova, G.~V.\ 
  2002, mnras, 333, 932 

\bibitem[Komissarov(1997)]{K97}
  Komissarov, S.~S.\ 
  1997, Phys. Lett. A, 232, 435 

\bibitem[Komissarov(1999)]{K99}
  Komissarov, S.~S.\ 
  1999, mnras, 303, 343, (KO) 

\bibitem[K{\"o}nigl \& Granot(2002)]{KG02} 
  K{\"o}nigl, A., \& Granot, J.\ 
  2002, ApJ, 574, 134 

\bibitem[Leismann et al.(2005)]{LAAM05} 
  Leismann, T., Ant{\' o}n, L., Aloy, M.~A., 
  M{\" u}ller, E., Mart{\'{\i}}, J.~M., Miralles, 
  J.~A., \& Ib{\' a}{\~ n}ez, J.~M.\ 
  2005, Astronomy \& Astrophysics, 436, 503 

\bibitem[Li (2005)]{Li05}
  Li S., 
  2005, J. Comput. Phys., 344-357

\bibitem[Londrillo \& Del Zanna(2000)]{LdZ00} 
  Londrillo, P., \& Del Zanna, L.\ 
  2000, ApJ, 530, 508 

\bibitem[Londrillo \& Del Zanna(2004)]{LdZ04} 
  Londrillo, P., \& Del Zanna, L.\ 
  2005, Journal of Computational Physics, 195, 17 

\bibitem[Macchetto(1999)]{Macchetto99} 
  Macchetto, F.~D.\ 
  1999, Astrophysics ans Space Science, 269, 269 

\bibitem[MacFadyen \& Woosley(1999)]{McFW99} 
  MacFadyen, A.~I., \& Woosley, S.~E.\ 
  1999, ApJ, 524, 262 

\bibitem[Mart{\'{\i}} \& M{\" u}ller(2003)]{MM03}
  Mart{\'{\i}}, J.~M.~\& M{\" u}ller, E.\ 
  2003, Living Reviews in Relativity, 6, 7 

\bibitem[Meier(2003)]{Meier03} 
  Meier, D.~L.\ 
  2003, New Astronomy Review, 47, 667 

\bibitem[Meszaros \& Rees(1994)]{MR94} 
  Meszaros, P., \& Rees, M.~J.\ 
  1994, MNRAS, 269, L41 

\bibitem[McKinney \& Gammie(2004)]{McKG04} 
  McKinney, J.~C., \& Gammie, C.~F.\ 
  2004, ApJ, 611, 977 

\bibitem[McKinney(2005)]{McK05} 
  McKinney, J.~C.\ 
  2005, ApJL, 630, L5 

\bibitem[Mignone et al.(2005a)]{MPB05}
  Mignone, A., Plewa, T., and Bodo, G.\
  2005, ApJS, 160, 199

\bibitem[Mignone \& Bodo (2005)]{MB05}
  Mignone, A., and Bodo, G.\
  2005, MNRAS, 364 (1), 126 (Paper I)

\bibitem[Mignone et al.(2005)]{MMB05}
  Mignone, A., Massaglia, S., and Bodo, G.\
  2005, Space Sci. Rev., in press

\bibitem[Miyoshi \& Kusano(2005)]{MK05} 
  Miyoshi, T., \& Kusano, K.\ 
  2005, Journal of Computational Physics, 208, 315 

\bibitem[Noh(1987)]{Noh87}
  Noh, W.F.\
  1987, J. Comput. Phys., 72,78

\bibitem[Romero et al.(2005)]{RMPIM05}
   Romero, R., Marti, J.~M., Pons, J.~A., Ibanez, J.~M., \& Miralles, J.~A.\ 
   2005, ArXiv Astrophysics e-prints, arXiv:astro-ph/0506527   

\bibitem[Rosswog et al.(2003)]{RRD03} 
  Rosswog, S., Ramirez-Ruiz, E., \& Davies, M.~B.\ 
  2003, MNRAS, 345, 1077 

\bibitem[Saltzman(1994)]{Saltzman94}
  Saltzman, J.\
  1994, J. Comput. Phys., 115, 153

\bibitem[Shapiro(2005)]{Shapiro05} 
  Shapiro, S.~L.\ 
  2005, ApJ, 620, 59 

\bibitem[T{\'{o}}th(1997)]{Toth00}
  T{\'{o}}th, G.\ 
  2000, J. Comput. Phys., 161, 605

\bibitem[Toro(1997)]{Toro97}
  Toro, E.~F.\ 
  1997, Riemann Solvers and Numerical Methods for Fluid Dynamics,
  Springer-Verlag, Berlin

\bibitem[Toro et al.(1994)]{TSS94}
  Toro, E.~F., Spruce, M., and Speares, W.\ 
  1994, Shock Waves, 4, 25

\bibitem[van Leer(1977)]{vLeer77} 
  van Leer, B.\ 
  1977, J. Computat. Phys., 23, 263 

\bibitem[van Putten(1993)]{vP93}
  van Putten, M.H.P.M.\
  1993, J. Comput. Phys., 105, 339 

\bibitem[Varni{\`e}re et al.(2002)]{VRT02} 
  Varni{\`e}re, P., Rodriguez, J., \& Tagger, M.\ 
  2002, Astronomy and Astrophyiscs, 387, 497 

\end{thebibliography}
\end{document}